# A visual analysis of the process of process modeling[1]

Jan Claes[1,2], Irene Vanderfeesten[2], Jakob Pinggera[3], Hajo A. Reijers[2], Barbara Weber[3] and Geert Poels[1]

[1] Ghent University, Belgium
{jan.claes, geert.poels}@ugent.be

[2] Eindhoven University of Technology, The Netherlands
{i.t.p.vanderfeesten, h.a.reijers}@tue.nl

[3] University of Innsbruck, Austria
{jakob.pinggera, barbara.weber}@uibk.ac.at

**Abstract**. The construction of business process models has become an important requisite in the analysis and optimization of processes. The success of the analysis and optimization efforts heavily depends on the quality of the models. Therefore, a research domain emerged that studies the process of process modeling. This paper contributes to this research by presenting a way of visualizing the different steps a modeler undertakes to construct a process model, in a so-called PPMChart. The graphical representation lowers the cognitive efforts to discover properties of the modeling process, which facilitates the research and the development of theory, training and tool support for improving model quality. The paper contains an extensive overview of applications of the tool that demonstrate its usefulness for research and practice and discusses the observations from the visualization in relation to other work. The visualization was evaluated through a qualitative study that confirmed its usefulness and added value compared to the Dotted Chart on which the visualization was inspired.

# 1 Introduction

In the quest for knowledge about how to make process models of high quality, recent research focus is shifting from studying the quality of process models to studying the process of process modeling itself (PPM). The PPM is a phase in the process model development lifecycle where the mental view of the modeler on the process is formalized into a graphical process representation (Hoppenbrouwers, Proper, & Van der Weide, 2005) (see e.g. Fig. 2). It encompasses the course of action taken by the modeler to design/construct a (business) process model consisting of start and end event(s), activities, gateways, edges, etc. Such a process model artifact is created by a stepwise design process; e.g., first putting a start event on the canvas, then an activity, then an arc connecting the start event and the activity, etc. The aim of PPM research is mainly to determine the characteristics of the process of process modeling that have a positive impact on process model quality. It is not about *what* is a good process model, but *how* can a good process model be created.

In order to be able to study current (implicit) modeling approaches of various modelers, since 2010 various datasets with data about such construction activities were collected in a series of observational modeling sessions executed at different universities in Europe (Pinggera, Furtner, et al., 2013; Weber, Pinggera, Zugal, & Wild, 2010; Weidlich et al., 2010)[2]. These data are being used in order to examine the PPM, mainly from a control flow perspective. The goal of the study presented in this paper is to discover characteristics of various model constructions (i.e., instances of the PPM). Ultimately, these exemplar cases can be used to examine if more generic modeling patterns exist (e.g., Pinggera, Soffer, et al., 2013). The patterns can then be studied further to evaluate if certain modeling patterns impact process model quality in a positive way: i.e., the search for best practices that can be generalized in empirically validated process modeling guidelines and tool support (e.g., Claes et al., 2012).

This paper describes the design of a tool to recognize and analyze these patterns in a cognitive effective way. The cognitive fit theory (CFT) states that a certain cognitive task can be optimally performed if the task material is represented appropriately (Vessey & Galletta, 1991). According to this theory, the proper instrument to discover relationships in datasets, are diagrams (Larkin & Simon, 1987). In other words, a visual representation of the data is believed to be a means to improve the efficacy of the cognitive task to discover patterns in the data

[2] An overview of these modeling sessions can be consulted at http://bpm.q-e.at/experiments.

(Fekete, Van Wijk, Stasko, & North, 2008). Furthermore, humans excel in visual pattern recognition (Baird, Coates, & Fateman, 2003). Therefore, a visualization was designed to support researchers, practitioners and tool developers to get insights in the PPM to develop theory, training and tools for improving process model quality.

For the visualization described in this paper, inspiration was drawn from the process mining research field (Van der Aalst, 2011). Process mining techniques make use of historical data of various process executions (i.e., process instances) to graphically represent and analyze a particular process (Weijters & Van der Aalst, 2001). The PPM is a typical type of process that also can be analyzed with process mining techniques. More specific, the visualization described in this paper was based on the Dotted Chart (Song & Van der Aalst, 2007), which represents every recorded event of the different process instances in one diagram in such a way that patterns across multiple instances can graphically be discovered and at the same time one can zoom in on details about the events of a single instance. However, the Dotted Chart does not support the analysis of the PPM optimally. Hence, this paper presents a modified implementation, i.e., the PPMChart.

This paper extends the work in (Claes et al., 2013) in several ways. The previous version of the PPMChart stayed close to the Dotted Chart Analysis plug-in in ProM[3] (Van Dongen, De Medeiros, Verbeek, Weijters, & Van der Aalst, 2005) in order to not confuse users that were familiar with that plug-in. In contrast, the improved version in this paper was fully redesigned. Every property of the chart and configuration option of the tool was evaluated against cognitive principles to reassure cognitive effectiveness of the visualization and the tool. The visualization is then applied in different contexts, which results in an extensive list of 13 observations of which initial insights are discussed and that are linked to empirically tested hypotheses. A qualitative study shows that the visualization succeeds in its goal to be useful for the study of the PPM, with a higher cognitive effectiveness than the Dotted Chart.

The discovery of confirmed (causal) relations between the PPM and the quality of the resulting process model and the knowledge about the circumstances needed to take optimal advantage of these relations, can enable the efficient exploration of ways to help improve process model quality in general. Modelers can be trained to implement the optimal modeling strategy that maximizes their individual capacity of creating high quality models in a specific domain or situation. Tools can be complemented with the developed knowledge to excel in supporting modelers to increase model quality. The PPMChart is a cognitive effective instrument to explore the data in order to build the necessary knowledge for the development of such training and tool support.

This paper reports on the development and application of the PPMChart visualization. The design science research method of Peffers, et al. is used to structure the paper (Peffers, Tuunanen, Rothenberger, & Chatterjee, 2007). The *problem description* is described in Section 2. The details of the *developed* graphical representation and the implementation are the subject of Section 3. The extensive overview of applications exhibited in Section 4 serves as a *demonstration* of the usefulness of the visualization in concrete analyses. Section 5 presents the result of a qualitative *evaluation*. Limitations and implications are discussed in Section 6. Section 7 describes related work. Finally, Section 8 summarizes the paper and discusses the need for and the value of this work.

# 2 Motivation

The goal of the research presented in this paper is to develop a visualization that can assist the study of the process of process modeling (PPM) in a cognitive effective fashion. Cognitive effectiveness is defined as the speed, ease and accuracy with which a representation can be processed by the human mind (Larkin & Simon, 1987). The inspiration for the PPMChart visualization was drawn from the Dotted Chart (Song & Van der Aalst, 2007). It was perceived to have an optimal balance between representing information about the structure of the overall process and the timing and relation of individual events. First, the Dotted Chart is presented in Section 2.1. Next, Section 2.2 evaluates the Dotted Chart as a solution for graphical analysis of the PPM. It can be concluded that the Dotted Chart in its current implementation is inadequate for studying the PPM in a cognitive effective way and that an adapted visualization is needed.

## *2.1 Dotted Chart*

The Dotted Chart visualization displays the events of the instances of a process as colored dots on timelines. Each timeline corresponds with one particular execution instance of the analyzed process and each colored dot on the timeline corresponds with a specific event for that process instance. The color of the dot indicates which event happened, while the position of the dot on the timeline represents the time when the event occurred (see Fig. 1a). It can be observed from Fig. 1a that it is difficult to analyze a Dotted Chart that contains information of all recorded PPM instances. For example, it is not possible to know which dots represent different events on the same model element (e.g., creation, movement, renaming of a *particular* activity) without manually

[3] The only implementation of the Dotted Chart we are aware of, is the Dotted Chart Analysis plug-in in the process mining framework ProM

investigating attributes in the event log (i.e., without leaving the plug-in). Therefore, one could focus on a single PPM instance if the event log is split into multiple event logs (one for each instance). The events in these event logs can then be grouped per model element (rather than per PPM instance). In this case, a Dotted Chart taking one such event log as input represents the operations of only one PPM instance (see e.g. Fig. 1b). We conclude, that it is possible to use the Dotted Chart for the visual study of the PPM. In the next section, we discuss the cognitive effectiveness of a visualization in general and of the Dotted Chart in particular.

(a) Full event log: multiple PPM instances (b) Transformed partial event log: only one PPM instance

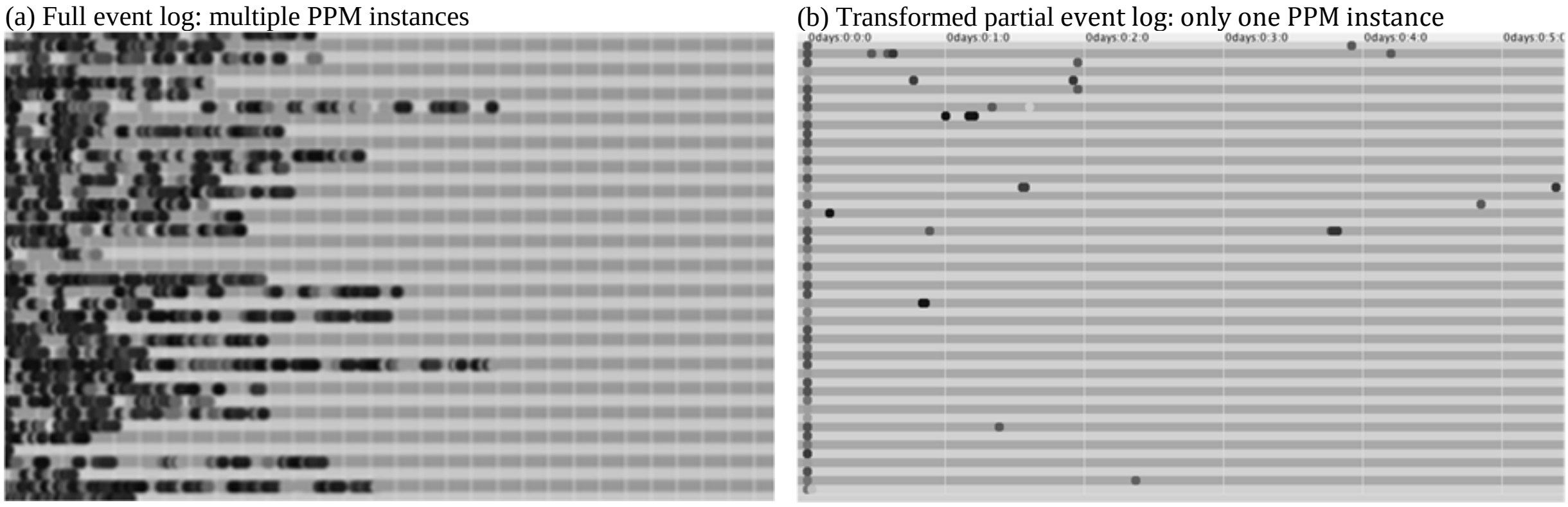

Fig. 1. Example of a Dotted Chart for the full event log with multiple PPM instances and for an event log containing events of only one PPM instance (each line representing the operations on a different process model element).

### 2.2 *Cognitive effectiveness of a visualization*

Before the PPMChart visualization is described in more detail in Section 3, this section explains what is meant with a '*cognitive effective visualization'*. Moody collected nine concrete principles that can be applied to improve the cognitive quality of visual notations, which are presented and discussed below (Moody, 2009). The design principles are also used to evaluate the Dotted Chart (when used on an event log containing a single PPM instance). The conclusion is that even with a limited amount of data in the event log, the cognitive effectiveness of the Dotted Chart can be substantially improved for the study of the PPM.

#### 2.2.1 Visual expressiveness

A graphical representation is visually expressive if it makes optimal use of the different graphical variables on which symbols can differ (Moody, 2009). Eight graphical variables are defined: shape, size, color, brightness, orientation, texture, horizontal position and vertical position (Bertin, 2010). Moreover, some graphical variables have a stronger impact on the cognitive load for interpreting the diagrams. Color is considered the most effective graphical variable (Lohse, 1993; Treisman, 1982; Winn, 1993), although in some situations it causes problems with visual perception (e.g., color blindness, black-and-white printers) (Moody, 2009). Therefore, other graphical variables should be used in combination with color.
The visual expressiveness of the Dotted Chart is rather low. With default settings, it makes no use of shape, size, brightness, orientation or texture. The various dots only differ in color and position. This can easily be improved by introducing the use of different shapes, brightness, size and/or texture of dots.

#### 2.2.2 Perceptual discriminability

Perceptual discriminability advocates that symbols are clearly distinguishable and that the more two concepts differ from each other, the more the corresponding symbols should differ (Winn, 1990). In this context, the *visual distance* is determined by the number of graphical variables for which two symbols have different values and by the size of these differences (Moody, 2009).
The perceptual discriminability of the Dotted Chart is also rather low. The colors of the dots are assigned randomly to the event classes that are present in the event log, which means that two dots with similar colors (e.g. blue and violet), do not necessarily represent similar events. Therefore, to increase perceptual discriminability, it is proposed (i) to assign fixed colors to fixed operations[4], (ii) to choose colors in such a way that similar events get similar colors, and (iii) to introduce other graphical variables such as shape and brightness in order to distinguish more easily between different dots.

#### 2.2.3 Graphic economy

There is a limit on the amount of different *values* for each graphical variable to assure that these design principles increase cognitive effectiveness (Nordbotten & Crosby, 1999). The span of absolute judgment, the amount of distinct perceptual values for each graphical variable, is estimated at seven (Miller, 1956). Furthermore, the amount of different objects that can be distinguished at a glance, i.e., the span of attention, is estimated at six objects (Miller, 1956).
It can be observed from the examples in Fig. 1 that the Dotted Chart violates the graphic economy principle. There are too many colors, which indeed hampers interpretation of the chart. In order to increase graphic economy, we propose to reduce the number of used colors, which can be easily compensated with the introduction of color shades (i.e., brightness) as discriminating variable.

[4] Note that this is only possible when there is a fixed, known set of possible occurring operations, which is the case in the study of the PPM.

### 2.2.4 Dual coding

While a graphical representation is a lot more effective than a textual representation (for processing information), the combination of both results in an even higher cognitive effectiveness (Paivio, 1990).
The Dotted Chart is divided in configurable time intervals of which *start and end date and time are textually displayed* on top of the chart. Each line also represents the *instance identifier at the beginning of the line*[5]. Information on selected dots is displayed in *tooltip text* when the mouse hoovers over the selected dot(s). We conclude that the Dotted Chart makes sufficient use of this design principle.

### 2.2.5 Semiotic clarity

Semiotic clarity means that every concept is represented by exactly one symbol and every symbol represents exactly one concept (Goodman, 1968).
Every event class of the event log is represented in a Dotted Chart as a dot with a unique color. Moreover, each dot has a unique position. Therefore, the Dotted Chart has maximal semiotic clarity. Nevertheless, it can sometimes be appropriate to introduce symbol deficit (i.e., use the same symbol for different concepts) to increase graphic economy (Moody, 2009).

### 2.2.6 Semantic transparency

A visualization has optimal semantic transparency if a novice would be able to guess the meaning of each symbol (Moody, 2009). This can be achieved through natural mappings: i.e., "*taking advantage of physical analogies and cultural standards*" (Norman, 2002, p. 23).
The Dotted Chart makes use of only three graphical variables: color, horizontal position and vertical position. When utilizing Dotted Charts for presenting the PPM, only one of these three variables is semantically transparent: the horizontal position indicates the timing of the corresponding event. The meaning of each color is not transparent to the reader of the chart (but can be derived from the color legend in a separate tab page). Similarly, it is difficult to know which horizontal line corresponds to which instance in the event log. For this, the reader first needs to zoom in on the chart to reveal the instance identifier displayed on each line[6]. Then, the reader can use this identifier to look up the necessary information in the attributes in the event log.

### 2.2.7 Complexity management

To cope with perceptual and cognitive limits (Miller, 1956; Novak, 2002), it is encouraged to reduce complexity by modularization (divide the diagram in smaller subsystems) and hierarchical structuring (make separate diagrams of the same information at different levels of abstraction) (Moody, 2009; R. Weber, 1997).
We suggest to only represent one PPM instance at a time in each chart. This way, the timelines can represent the different model elements (as in Fig. 1b), rather than different instances of the PPM (as in Fig. 1a). This can be achieved by splitting the event log in multiple event logs, each containing information about only one PPM instance. Further, it is possible in the Dotted Chart to customize the chart to abstract from certain differences or tailor the view to a certain analysis. All the same, complexity can further be reduced by filtering the displayed information. This can optimally be managed from within the plug-in, rather than at event log level (which is currently the only option in the Dotted Chart implementation).

### 2.2.8 Cognitive integration

When different diagrams are used, explicit mechanisms should exist to support the integration of these diagrams (Hahn & Kim, 1999; Kim, Hahn, & Hahn, 2000).
When different PPM instances are represented by different charts (rather than aggregating them in only one chart), the need for cognitive integration mechanisms emerges. A first step towards cognitive integration might be a *uniform color, shade and shape coding and time scaling* between different charts. Without uniform coding (as in the Dotted Chart), comparing or combining information of multiple charts is not cognitive efficient.

---

[5] For readers that are familiar with process mining: this is the trace identifier in the event log.
[6] Due to a misalignment, this is not visible in a typical Dotted Chart without zooming in (this might be an unintentional bug).

### 2.2.9 Cognitive fit

The optimal representation of data depends on the task it supports and on the user of the visualization. For the same representation, the cognitive load is greater for novices than for experts (experts in working with that particular visualization) (Vessey & Galletta, 1991). Similarly, depending on the particular analysis the user needs the representation to support, a different view on the data is desired.
The Dotted Chart has various *configuration options*, which facilitate to tailor the appearance of the chart according to the needs of the analysis. However, a mechanism to filter the information that is displayed in the chart is cumbersome, because it involves leaving the plug-in, using a filter plug-in of the framework on the event log, regenerating a chart from the filtered log and reconfiguring the chart to customize the view.

### 2.2.10 Conclusion

The Dotted Chart can be used in its current form to study the PPM. To maximize complexity management, it seems appropriate to represent only one PPM instance at a time. However, according to the presented design principles for visual notations, cognitive effectiveness of the Dotted Chart can be substantially improved in the specific context of this research (even if an event log containing data about a single PPM instance is provided to the plug-in). For 7 out of 9 principles concrete suggestions are formulated, which formed the basis for the extension of the Dotted Chart presented in this paper. Note that every suggested improvement is specific for the PPM and could therefore not be incorporated in the original Dotted Chart plug-in. Because the improvements are specific for the context of the study of the PPM, it was decided to call the improved charts PPMCharts.

# 3 The PPMChart visualization

The PPMChart visualization graphically represents data of the process of process modeling (PPM) (see Section 3.1). The visualization uses timelines on which colored dots are associated with positions that correspond to the time when the corresponding PPM operations are recorded for the process model (see Section 3.2). The Dotted Chart Analysis plug-in was adapted and extended (see Section 3.3), which resulted in a new plug-in in the ProM tool that produces the PPMCharts (see Section 3.4).

## *3.1 Data requirements*

The PPM is a human endeavor in which a modeler constructs a process model by drawing model elements such as activities, events, gateways, and edges on a canvas (see Fig. 2). In order to be able to represent a PPM instance with the PPMChart visualization, data of the PPM instance need to be collected at a specific level (see Section 3.4.1). Therefore, it is convenient if a modeling tool with logging functionality is used for the construction of the process model. The PPMChart implementation is created under the assumption that it is possible to record data on every modeling operation on the canvas (e.g., create start event, create activity, move activity). Besides the *name* of each operation, the visualization needs two more attributes: the *identifier* of the model element on which the operation was performed and the *timestamp* of the execution of the operation. Possible other recorded attributes (such as the position of a model element on the canvas) are ignored by the visualization.

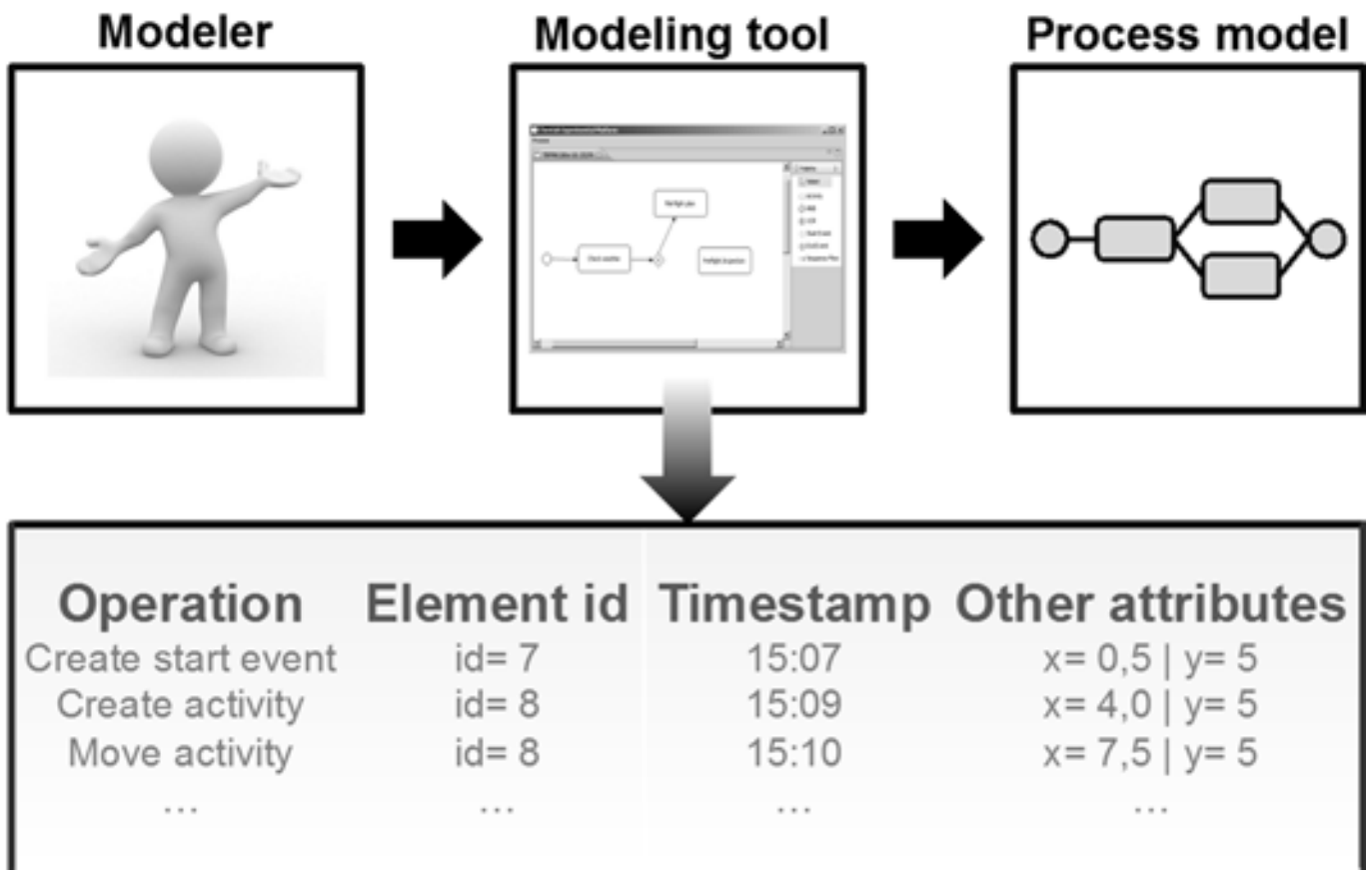

Fig. 2. The process of process modeling and the attributes of the captured data.

## *3.2 Visualization with PPMChart*

The collected data about consecutive operations in a PPM instance are used to construct a PPMChart (see Fig. 3). The horizontal axis represents a time interval of one hour by default. Vertically, each line represents one element of the process model, as it was present during modeling. The model element identifier is displayed at the beginning of the line. Each colored dot on the line represents one operation performed on the element.

- The color of the dot corresponds with the *type of operation*: create (green), move (blue), delete (red) and (re)name (orange).
- The color shade and shape of the dot corresponds with the *type of model element*: activity (bright, box), event (very light, circle), gateway (dark, diamond) and edge (light, triangle).
- The position of the dot on the timeline corresponds with the *time* at which the operation was executed.

The user can configure the order in which the timelines are presented. The default order of lines corresponds with the logical order from start event to end event of the elements in the process model (see Section 3.4.4).

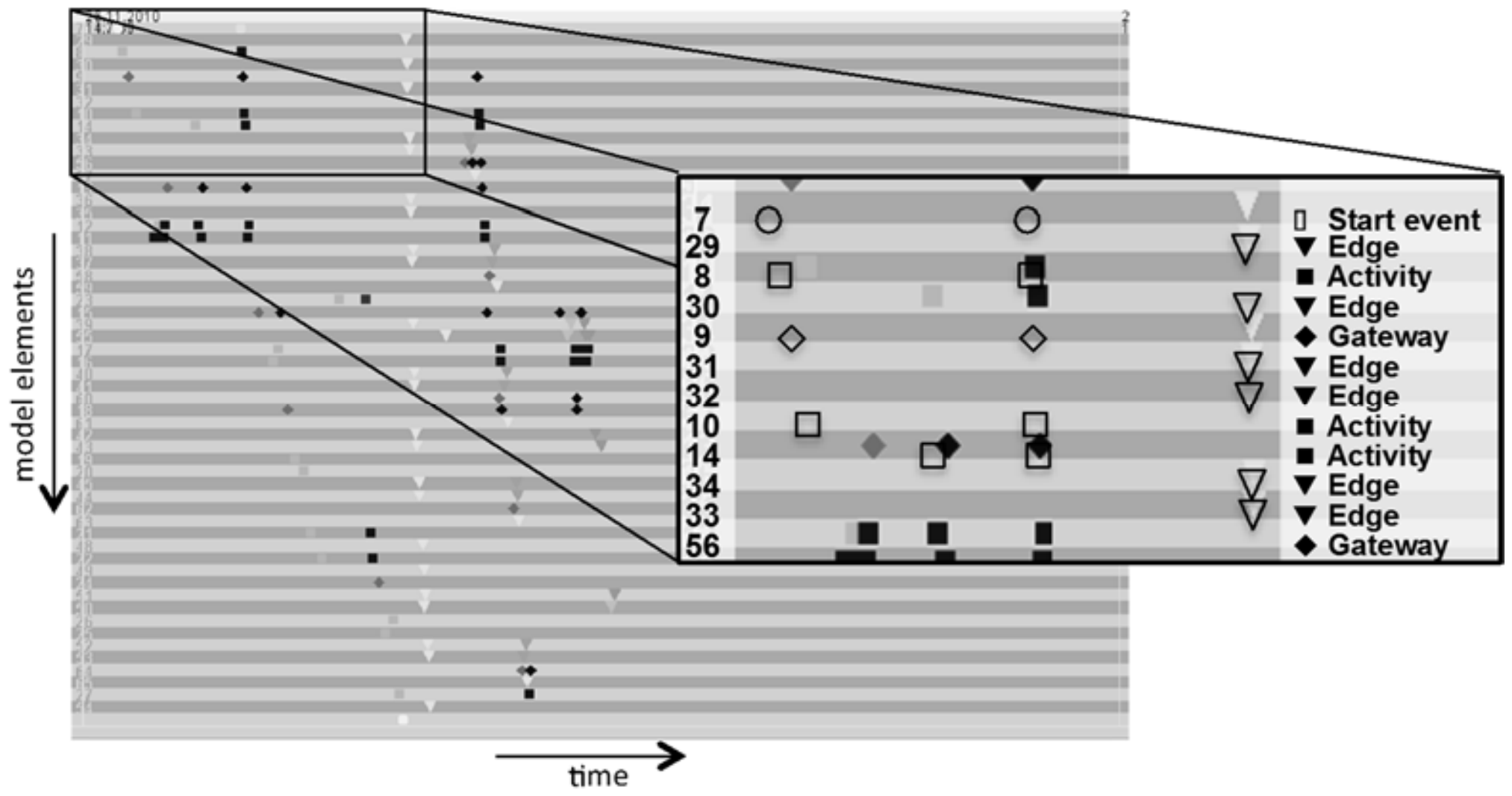

Fig. 3. Visualization of the events in the construction of one model by one modeler.[7] (85)

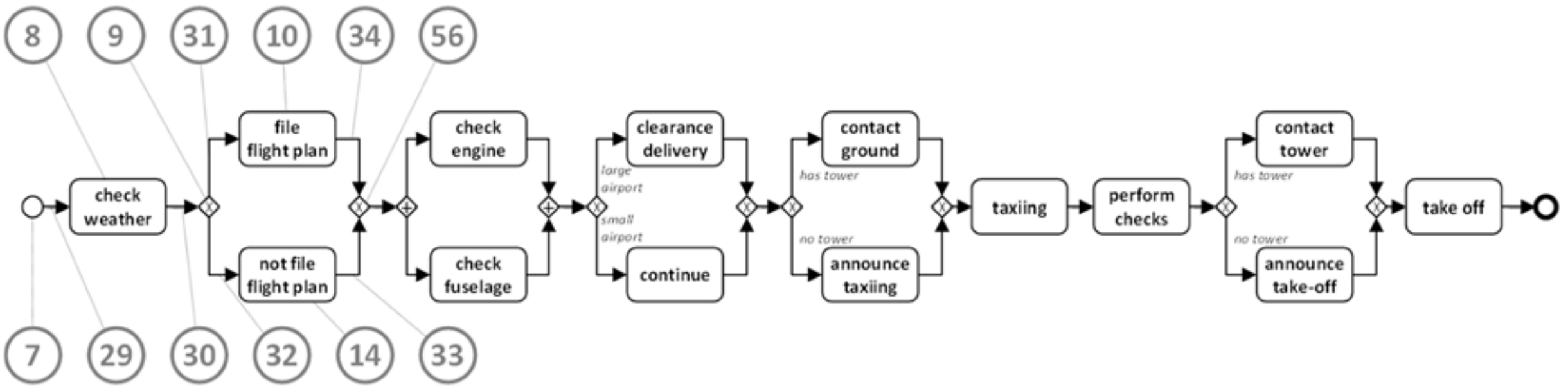

Fig. 4. Process model in BPMN notation as result of the modeling process in Fig. 3.[8]
(Numbered circles indicate the model element id for comparison with highlight of Fig. 3)

Fig. 3 shows the different operations in the creation of the process model represented in Fig. 4. The highlighted rectangle in Fig. 3 displays the first operations on the left part of the model in Fig. 4 (from start event to first XOR join gateway). On the first line, one can observe the creation of the start event (very light green circular dot). More to the right a very light blue circular dot represents a movement of that start event. The second line shows a light green triangle at the right that corresponds with the creation of the edge that connects the start event and the first activity. Within the highlighted rectangle, different other light green triangles can be discovered on different lines. These dots represent the creation of other edges in the process model. The third line contains a bright green square shaped dot. This is the creation of an activity. More to the right, a blue square shaped dot indicates that this activity is moved later on. If one focuses on all colored dots in the rectangular selection, one can conclude that the creation of start event, activities and split gateway in the highlighted section is followed by an almost simultaneous movement of these model elements (vertical pattern of blue dots). Only later (i.e., more at the right), the edges that connect these model elements were created (light green triangles).

### *3.3 Differences with Dotted Chart*

The PPMChart visualization differs from the classical Dotted Chart mainly in four ways:

---

[7] High resolution graphs in color of all figures in this paper are available from http://www.janclaes.info/papers/PPMISeB.

[8] Download an animation showing how the PPMChart evolves during the model construction at http://www.janclaes.info/papers/PPMISeB.

(i) In contrast to the typical use of a Dotted Chart, the PPMChart displays information about only one process instance. Each timeline represents a particular *model element* of the constructed process model (i.e., activity, gateway, edge, etc.). The events in the PPM, represented by the colored dots, are the *operations* (i.e., creation, movement, deletion, etc.) performed on the particular model element represented by the timeline. See Section 3.4.1.

(ii) The PPM has a fixed set of possible operations and therefore, in a PPMChart, these operations are mapped on fixed default colors, color shades and shapes, which eases the visual comparison of different charts. For the same reason, every PPMChart shows initially information of the same timespan (i.e., one hour). Nevertheless, the option to change the color (shade) and shape coding of the dots and to zoom in or zoom out to influence the displayed timespan still exists. See Sections 3.4.2 and 3.4.3.

(iii) For different graphical pattern analyses, the Dotted Chart can be sorted according to several sort options. Two sort options are added in the PPMChart implementation, which facilitate the study of the PPM with the charts. See Section 3.4.4.

(iv) The option to filter certain operations (individual dots) or model elements (individual timelines) enables the analyst to take different views at different abstraction levels on the data from within the PPMChart implementation. For example, one can focus on creation of model elements if other operation types are filtered out. See Section 3.4.5.

### *3.4 Tool support*

The *PPMChart Analysis* plug-in[9] (see Fig. 5) is an adapted version of the existing *Dotted Chart Analysis* plug-in in ProM. In the middle, the PPMChart is presented. At the left hand side, one can configure the view, and at the right one can filter the data. The previous version of this plug-in (Claes et al., 2013) remained closely to the Dotted Chart Analysis plug-in implementation in order to not confuse users that are familiar with that plug-in. The version described here was fully redesigned after feedback of various experts (i.e., participants to the BPM 2012 conference, international process modeling and visualization experts at various occasions). Every visualization property and tool setting was evaluated against principles of cognitive efficacy optimization.

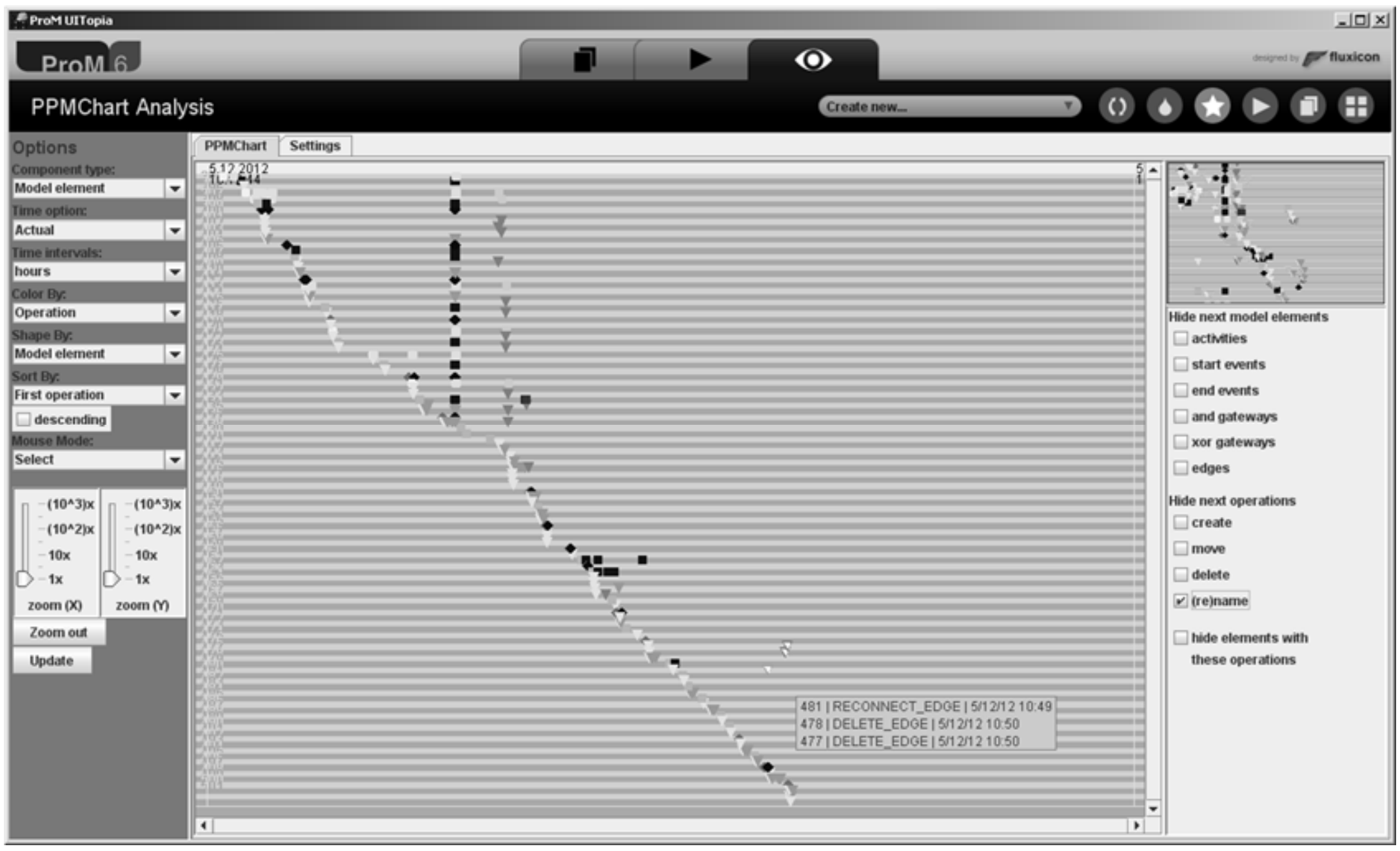

Fig. 5. Screenshot of the PPMChart window in ProM. (Model 2012-184)

This section describes the data format for the PPMChart plug-in and the configuration options that were added to the Dotted Chart Analysis plug-in. A more detailed description about every configuration option of the implementation can be found in Appendix A.

### 3.4.1 Required format of the recorded data

For the sake of *complexity management* and *cognitive fit*, it was decided to define a PPMChart as a visualization to display information of a single process modeling effort. Therefore, the tool requires as input an event log containing data about a single PPM instance, grouped in traces per model element. Section 3.1 already explained which data of that PPM instance are needed. The required format of these data is described in this section.

---

[9] The PPMChart Analysis plug-in for ProM 6 can be downloaded at http://www.janclaes.info/plugins.php.

In order to visualize the PPM, a fixed list was selected of possible operations in the construction of a process model (see Table 1). In our analyses and modeling sessions we build on a subset of the BPMN notation that can be used for the modeling. This subset was selected to correspond with the supported notation of our modeling tool Cheetah Experimental Platform (CEP) (Pinggera, Zugal, & Weber, 2010) and consists of six of the ten most used elements of BPMN according to (Zur Muehlen & Recker, 2008): start and end event, activity, XOR and AND gateway, and edge. It can be considered as the set of common constructs that are available in most process modeling languages (e.g., BPMN, UML Activity diagrams, Petri nets, Workflow nets, YAWL).

Besides creation of these model elements, the visualization also includes changes in the model. Activities, events and gateways can be moved over the canvas or deleted. Edges can be deleted or reconnected, an edge can be rerouted through creation, movement and deletion of edge bend points, and the label of an edge can be moved. Finally, activities and edges can be named or renamed. Note that for the remainder of the paper we assume only these modeling operations as part of the PPM (according to the recorded operations of the modeling tool), but our approach can easily be adapted for other modeling operation sets.

Table 1. Operations in the construction of a process model.

| Create | Move | Delete |
|---|---|---|
| CREATE_START_EVENT | MOVE_START_EVENT | DELETE_START_EVENT |
| CREATE_END_EVENT | MOVE_END_EVENT | DELETE_END_EVENT |
| CREATE_ACTIVITY | MOVE_ACTIVITY | DELETE_ACTIVITY |
| CREATE_XOR | MOVE_XOR | DELETE_XOR |
| CREATE_AND | MOVE_AND | DELETE_AND |
| CREATE_EDGE | MOVE_EDGE_LABEL | DELETE_EDGE |
| | RECONNECT_EDGE | |
| **Other** : NAME_ACTIVITY, RENAME_ACITIVTY, NAME_EDGE, RENAME_EDGE, CREATE_EDGE_BENDPOINT, MOVE_EDGE_BENDPOINT, DELETE_EDGE_BENDPOINT | | |

A plug-in for the well-known academic process mining framework ProM[10] was developed to facilitate the creation of PPMCharts. The input for most plug-ins in this tool is an event log. The xes file format for event logs for ProM is xml based and follows a certain hierarchical structure: a *process* consists of *traces* and each trace is a collection of *events*[11]. The process, traces and events can have attributes. In order to be able to create a PPMChart, the *events* in the log have to store information about operations on model elements and the events that correspond with operations on the same model element have to be bundled in one *trace*. The plug-in expects the *names* of events to correspond with the operations in Table 1. Further, each event should have an attribute that stores the *timestamp* of the execution of the event and an attribute *id* that matches with the name of its trace. It can be seen as the unique identifier of the model element.

### 3.4.2 Fixed default color, shade and shape coding

*Visual expressiveness* is increased by the addition of default shade and shape coding. In order to preserve *semantic transparency*, size, texture and orientation are currently not used as symbol discriminating factors. The selection of similar colors and shapes for similar operations improves *perceptual discriminability*. This introduces some redundant coding (i.e., the type of model element can be deduced from the color shade and from the shape) (Green & Swets, 1966), but increases perceptual pop-out (because almost each different operation has its unique color (shade), it is easy for the human brain to filter out specific operations/colors) (Quinlan, 2003; Treisman & Gelade, 1980).

The introduction of shades of colors reduces the amount of necessary colors, which adheres to the principle of *graphic economy*. The amount of used graphical variables raises (i.e., increased visual expressiveness), while the amount of *values* for each graphical variable can be limited to a lower number (i.e., graphic economy). Because the goal of the PPMChart is to reduce cognitive load and increase cognitive effectiveness for the study of data on the PPM, much importance was attached to this principle. The graphic economy of a representation with more than six different colors can be increased by (i) *increasing visual expressiveness*, (ii) *reducing semantic complexity*, and (iii) *introducing symbol deficit* (Moody, 2009). As a consequence, symbol deficit was introduced on operations on both types of events (i.e., start and end event) and both types of gateways (i.e., AND and XOR gateway). *Perceptual discriminability* advocates the choice for exactly these symbols to introduce symbol deficit, because they are most similar. However, the user can still change the color (shade) and shape of the dots such that both gateways and both events can be graphically distinguished.

[10] The ProM tool can be downloaded at http://www.promtools.org.

[11] The xes file format of ProM is described at http://www.xes-standard.org.

Furthermore, *semantic transparency* is increased by selecting logical colors. Shades of green represent create operations and shades of red represent delete operations. For move operations, the third primary color was selected (i.e., blue). The selected shapes are similar to the BPMN symbol for the element type (square for activities, circle for events, diamond for gateways). Lastly, *cognitive integration* is facilitated by selecting fixed default coding. This resulted in the fixed default color (shade) and shape coding as presented in Table 2. However, the user can still modify the color (shades) and shapes in the *Settings* tab.

Table 2. Default colors, shades and shapes of the PPMChart Analysis plug-in.

| Operation | Color | Model element | Shape | Color shade |
|---|---|---|---|---|
| Create | Green | Activity | Rectangle | Bright |
| Move | Blue | Event | Circle | Very light |
| Delete | Red | Gateway | Diamond | Dark |
| (Re)name | Orange | Edge | Triangle | Light |

### 3.4.3 Fixed default time interval

By default, the time interval represented by a PPMChart is fixed, which makes it easier to compare time related issues between different PPMCharts (i.e., facilitating *cognitive integration*). The length of this interval is set to an arbitrary value of one hour, but this interval can be modified with the use of the zoom (X) button.

### 3.4.4 Sort options

*Cognitive integration* also promotes the addition of two sort options that help the user to find a link between the PPMCharts and the corresponding process model elements: '*Distance from start*' and '*Create order from start*' (see Fig. 6). This subsection explains both additional sort options, as well as the option to sort by '*First operation*', which existed already in the Dotted Chart implementation and is used in two examples in this paper.

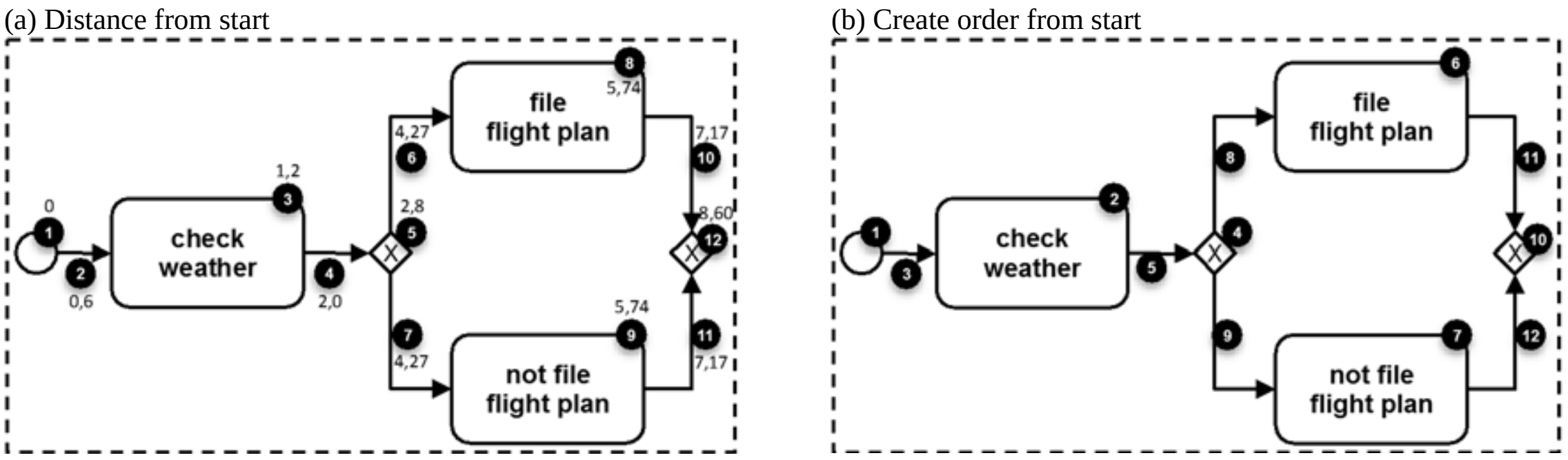

Fig. 6. Additional sort order options in the PPMChart implementation.

#### Distance from start

The *Distance from start* sort option was added to sort the timelines according to the processing order of the corresponding model elements from start event to end event. This can be observed in Fig. 6a, which represents the left part of the process model of Fig. 4 and where the order of model elements is indicated with black circle annotations. For elements in parallel paths the order is rather arbitrary (see technical details below).

To determine this order technically, we used the notion of *length* of an arc, *path*, path *distance* and *shortest path*. We define the *length* of an arc as the graphical distance between start and end point of the arc, regardless the actual routing of the arc through possible bend points. A *path* is a route between two elements in the process model and summarizes the consecutive model elements that are to be passed when traversing the model from A to B through nodes and directed arcs that connect these nodes. The *distance* of a path is the sum of the lengths of the arcs in the path. If an arc is the first or last element in the path, only half of its length is included in the distance. The *shortest path* between two nodes is the path that starts in one of the nodes and ends in the other node with the lowest distance. The *distance from start* orders each model element according to the distance from the shortest path from the start event to the model element. The distance from start for each model element in Fig. 6a is displayed right above or below the black circle annotation that indicates the order of the model elements. If two elements get the same distance value, the order of the related records in the log is preserved.

#### Create order from start

The *Create order from start* sort option was added under the assumption that in several modeling tools (e.g., ARIS, CEP) an edge connecting two nodes can only be created after the two nodes exist. Hence, it uses the same ordering mechanism as *Distance from start*, except that it puts each arc after the nodes it connects (see Fig. 6b). This sort option was added to resemble (our interpretation of) the *logical* order of creating the model elements in a process model from start event to end event (i.e., from left to right in the example).

Technically, while in the *Distance from start* ordering the value of the arc is the mean value of the nodes it connects, in the *Create order from start* ordering, it is the maximum value of the nodes it connects plus one.

**First operation**

The remainder of the article contains a lot of examples of PPMCharts. The majority of them apply one of the above sort options, but in two cases, it was useful to sort by *First operation*. This sort option orders the timelines according to the *actual* creation order of the model elements (the first operation of each element is its creation).

#### 3.4.5 Filter options

In order to raise *cognitive fit* and for *complexity management*, the PPMChart can be filtered from within the plug-in to show or hide dots on the charts. Model element types can be selected to hide all operations on all model elements of that type (e.g., hide operations on edges). The timelines representing these model elements remain in the chart in order to preserve the positioning of the remaining timelines, but the concerned dots on these timelines are hidden. Next, it is also possible to hide specific operations by selecting the operation type in the filter panel (e.g., hide (re)name operations). Every dot that represents an operation of the selected type will be hidden. Finally, if the checkbox at the bottom is selected, all operations of model elements for which an operation exist of the selected type are hidden (e.g., hide deleted elements). Every dot on the timelines that contain at least one dot representing an operation of the selected type is removed from the chart.

#### 3.4.6 Example

Fig. 5 above shows an example of the implementation. The lines are vertically sorted by '*First operation*'. On the first line a circle indicates that a start or end event was created first. The last line represents the last created model element. It appears to be an edge (triangle). At one moment in time a lot of elements were moved simultaneously (perfectly vertical blue line formed by blue dots of different shapes), and somewhat later it can be observed that a lot of elements were deleted right after each other (almost vertical red line). The deleted elements were mainly events and edges (circles and triangles). Immediately after these operations, a high number of edges were created or reconnected (light green and gray rectangles below the red vertical line).

# 4 Application

To demonstrate the usefulness of the PPMChart visualization, it was applied on the collected data for many process of process modeling (PPM) instances of students. A PPMChart was generated for every PPM instance in the dataset. First, a description of the data collection is provided (see Section 4.1). Next, a list of observations is presented, followed by their interpretation and a number of possible explanations, illustrated by examples from the dataset. Besides simple observations such as modeling time or amount of created model elements (see Section 4.2), more complex observations can be made from the study of patterns of operations (see Section 4.3). Furthermore, different charts can be compared to get additional insights (see Section 4.4).

The possible explanations were selected to optimally illustrate the usefulness of the PPMChart. It is not the purpose to be complete, but the focus is merely on demonstrating the usefulness of the visualization. A deeper understanding of the presented observations is discussed in Section 6. Note that the caption of each figure indicates which settings were used to produce the PPMChart in the plug-in.

### *4.1 Data collection*

The observations below are based on the data of two observational modeling sessions conducted at Eindhoven University of Technology. The participants were international master students of three different educational programs (i.e., Operations Management & Logistics, Innovation Management, and Business Information Systems), which attended a course in Business Process Management. Participation was voluntarily and the students could decide to stop at any time without handing in a solution. They firstly completed a tool tutorial to get familiar with CEP and the modeling language. In this tutorial the user was presented with an explanation and a short movie on how to perform a certain operation in the tool. Only when the participant successfully imitated the example, the tutorial continued. Next, based on a textual description they were asked to model the control flow of a certain business process. A survey was presented to collect additional information on the modelers (e.g., gender, age, familiarity with the case, etc.).

The first session was performed in November 2010 and 120 students participated. Each student constructed a process model in CEP for two cases (Pre-Flight and NFL case[12]). For the Pre-flight case, the modelers created models with 13 activities and used 120 recorded operations on average. The NFL models contained 9 activities and were constructed with 85 recorded operations on average. In December 2012, 117 students participated in the second modeling session. For this session, every student modeled only one case (Mortgage case[13]). Again, CEP was used to record the modeling operations. This session resulted in models of 27 activities and using 276 recorded operations on average. This indicates the mortgage case is more extensive than the other two cases.

[12] Both case descriptions can be downloaded from http://bpm.q-e.at/experiment/Pre-Flight.

### *4.2 Simple observations*

A visualization of high quality supports the discovery of surprising insights in highly complex data, but should also support the easy derivation of simple characteristics. For the sake of brevity, this section is limited to a brief presentation of four of such rather simple observations.

**Modeling speed**
***Observation***: The width of the part of the chart that contains dots differs between charts from different modelers that modeled the same case.
***Interpretation:*** Some modelers work faster than others.

**Modeling pauses**
***Observation***: Some charts have clear horizontal gaps: i.e., a non-trivial time interval where no line contains dots.
***Interpretation***: Some modelers pause their modeling operations at certain times.

**Amount of model elements**
***Observation***: The number of lines differs between charts from different modelers that modeled the same case.
***Interpretation***: Some modelers create more model elements.

**Amount of modeling operations**
***Observation***: The relative number of blue, red, and orange dots differs between charts from different modelers that modeled the same case.
***Interpretation***: Some modelers tend to move, delete, or rename elements relatively more than others.

### *4.3 Observations on patterns of operations on model elements*

This section discusses a number of observations that relate to patterns of operations on model elements. These can be derived from the amount and position of dots of a certain type.

#### 4.3.1 Patterns of delete operations

***Observation***: If the chart contains red dots, they are sometimes scattered around the chart (without a clear pattern), and sometimes a vertical line of red dots can be distinguished.
***Interpretation***: Some modelers delete model elements at various times, some modelers decide to delete a whole part of the model (i.e., multiple model elements) at the same time.

***Possible explanations***: In Fig. 7a the red dots are scattered over the PPMChart. Possibly, this means the modeler occasionally changed her/his mind about the content of the model. When the PPMChart shows a vertical line of delete operations as can be observed in Fig. 7b, the modeler threw away a whole part of the model at once. Possibly, the modeler wanted to start over and remodel that part or decided that that part was not necessary in the model (a closer inspection of the data about the operations after the deletion might reveal the exact cause).

(a) Model 2010-318 (b) Model 2010-213

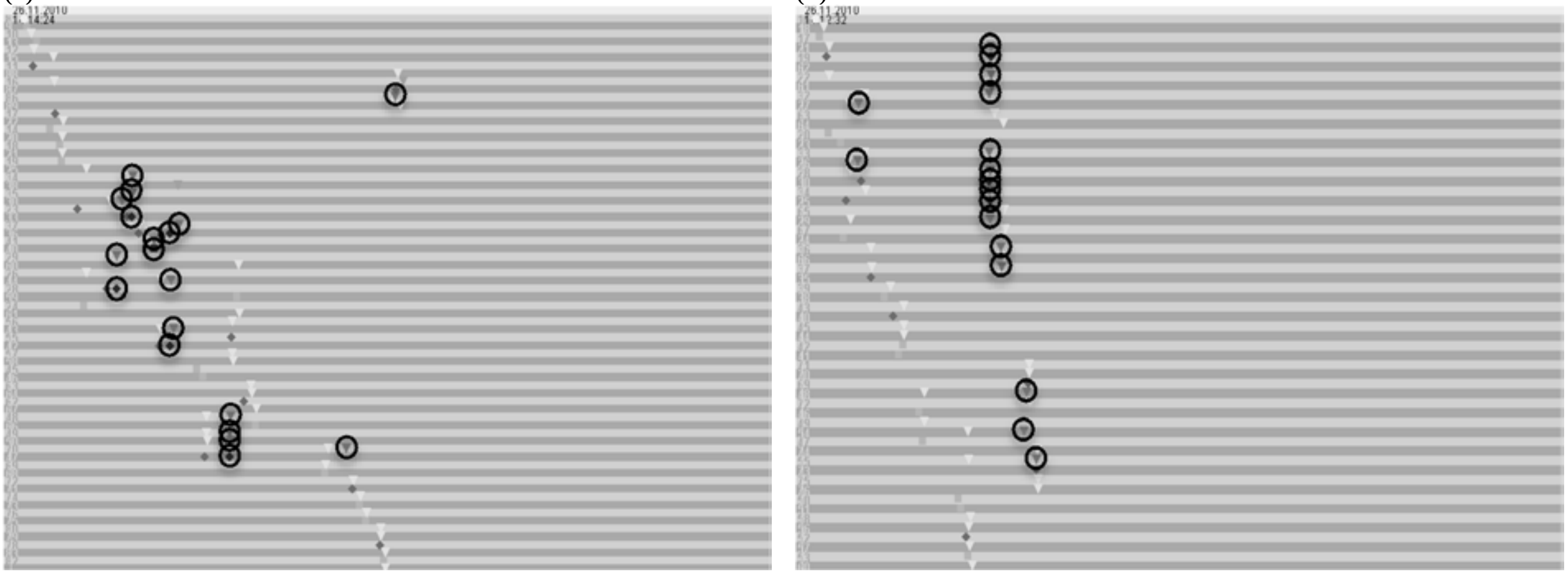

Fig. 7. Scattered or simultaneous delete operations.
(charts hide moves and renames)

[13] Case description can be downloaded from http://bpm.q-e.at/experiment/MortgageEindhoven.

### 4.3.2 Patterns of move operations

***Observation***: If the charts contains blue dots, sometimes they are positioned near to the green dot on the same line, sometimes they form a broad vertical line at the right, and sometimes the dots are scattered over the chart.
***Interpretation***: Some modelers hardly move any model elements, some modelers move the elements shortly after their creation, some modelers tend to move elements rather at the end of their modeling process, some modelers move model elements at various times.

***Possible explanations***: Some modelers do not move a lot of process model elements (see Fig. 8a). In Fig. 8b the movement pattern looks like a diagonal line. This means the modeler has moved elements only shortly after their creation and did never touch them again. Possibly, the modeler either has been pretty determined on the layout of the complete model, or the modeler did not bother to work on the layout of the model (the constructed process models might provide clarification). When the PPMChart shows a vertical line of blue dots at the right, the modeler has first created a number of model elements and then moved them simultaneously (see for example Fig. 8c). Possibly, the elements needed to be moved to make room for a new part of the model, or to layout the model better. Finally, when the chart contains the triangle-like movement pattern as presented in Fig. 8d, the modeler keeps moving elements that were created earlier in the modeling process. Possibly, this is caused by continuously layouting or creating space in the model, which might be derived from a closer review of the data.
Note that the modeler may also combine these patterns; for instance, a modeler moves the elements shortly after their creation, does not touch them again until she/he at the end starts moving elements around again to work on the overall layout of the resulting model. This is for example displayed in Fig. 8c where the PPMChart has, next to the vertical line of blue dots at the end, also a moving phase at the beginning of the modeling process.

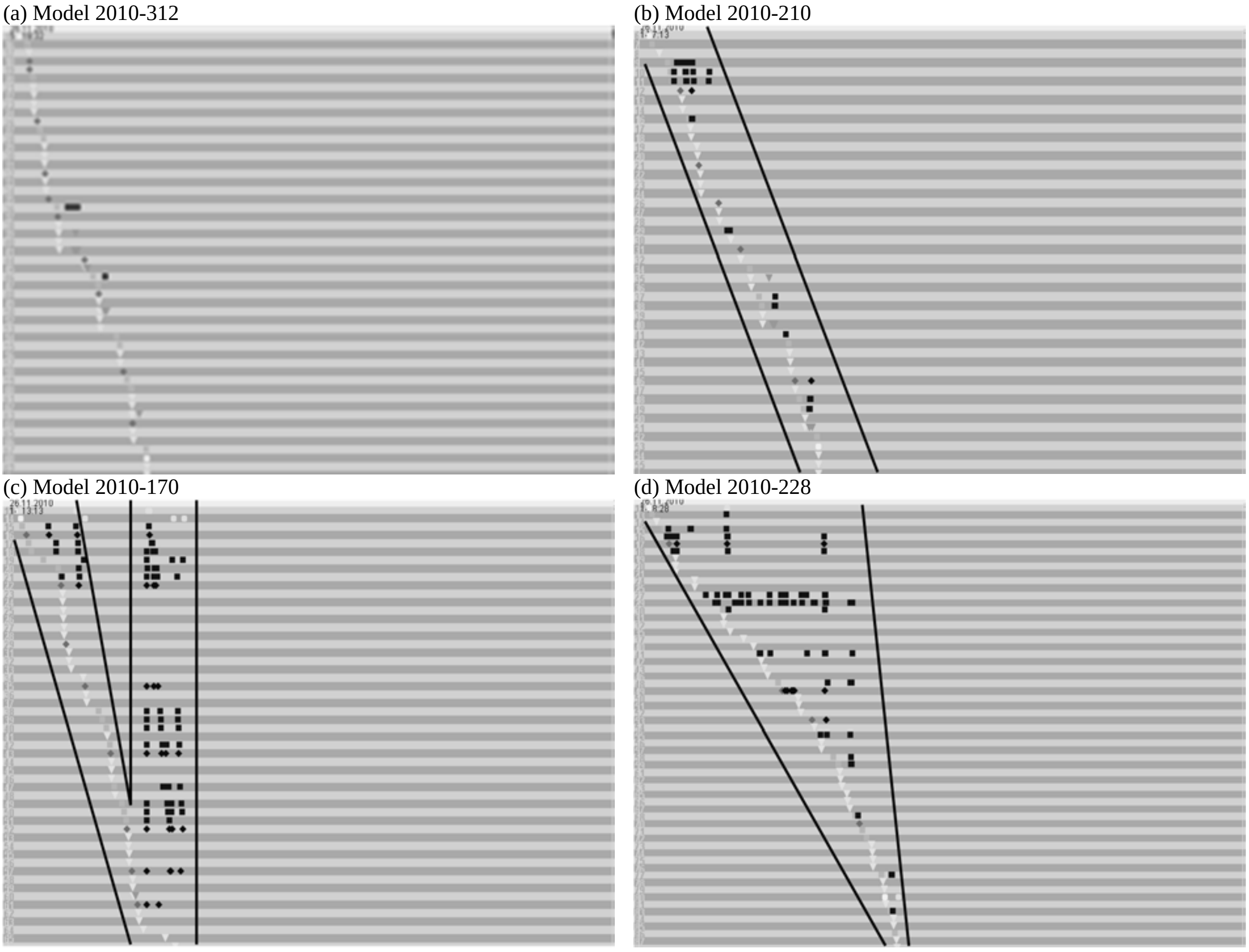

Fig. 8. Timing of movement operations: few (a), close to creation (b), at the end (c), scattered (d).
(charts are sorted by *First operation* and moves and renames are hidden)

### 4.3.3 Patterns of create operations: order of creating activities, gateways and edges

***Observation***: Some charts have non-crossing vertical lines formed by green, dark green and light green dots.
***Interpretation***: Some modelers delay creation of edges (and gateways) until all activities are put on the canvas.

***Possible explanations***: In some PPMCharts, activities are created first (green dots) followed by the edges (light green dots) (see Fig. 9a), while other PPMCharts show a more divers order of creating activities, gateways and edges (see Fig. 9b). Possibly, modelers either work aspect-oriented (i.e., they first focus on the content aspect by creating all activities in the model before connecting them with gateways and edges to fix the structure aspect), or flow-oriented (i.e., they first finish a logical part of the model by creating nodes and edges and then turn to another part of the model).

(a) Model 2010-367 (b) Model 2010-237

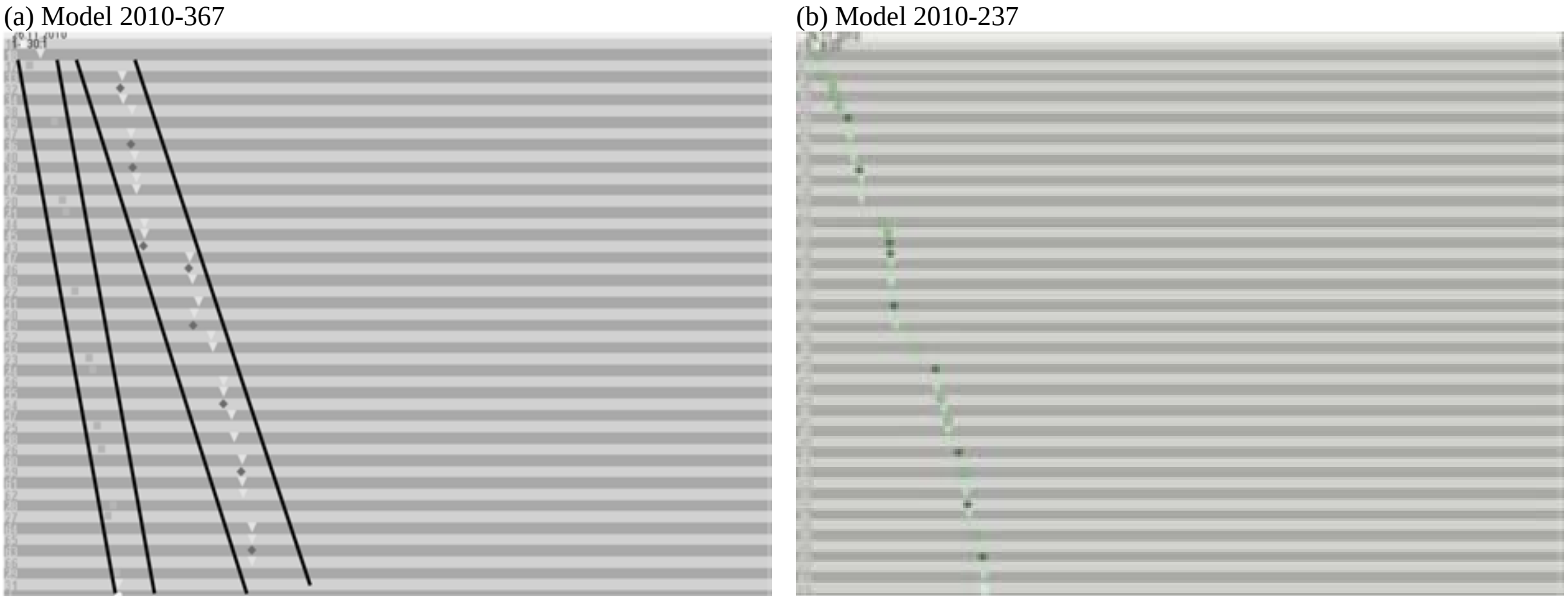

Fig. 9. Order of creation of activities, gateways and edges.
(charts hide moves, deletes and renames)

#### 4.3.4 Patterns of create operations: split and join gateways

***Observation:*** When sorted by first event, it can be noticed that in some charts dark green dots come (mostly) in pairs, while in other charts they are interchanged with bright green dots.
***Interpretation***: Some modelers create the join gateway right after the creation of the split gateway, while others create the join gateway later on.

***Possible explanations***: Fig. 10a presents a model where the creation of one gateway is mostly directly followed by the creation of another gateway. This modeler puts the split and join gateway right after each other on the modeling canvas, possibly to not forget to add the join gateway. He concentrates first on the correct structure of the model and then on the content (i.e., aspect-oriented). The modeler in Fig. 10b follows a more flow-oriented approach. The study of the chart leaves the impression that the modeler constructed blocks of the model more linearly from start to end event. The join gateways are only created when all intermediate activities of the block structure are already in place. Note, that for both modelers the overall modeling process is flow-oriented (constructing part after part), but only within the creation of model blocks a different approach is observed.

(a) Model 2010-106 (b) Model 2010-361

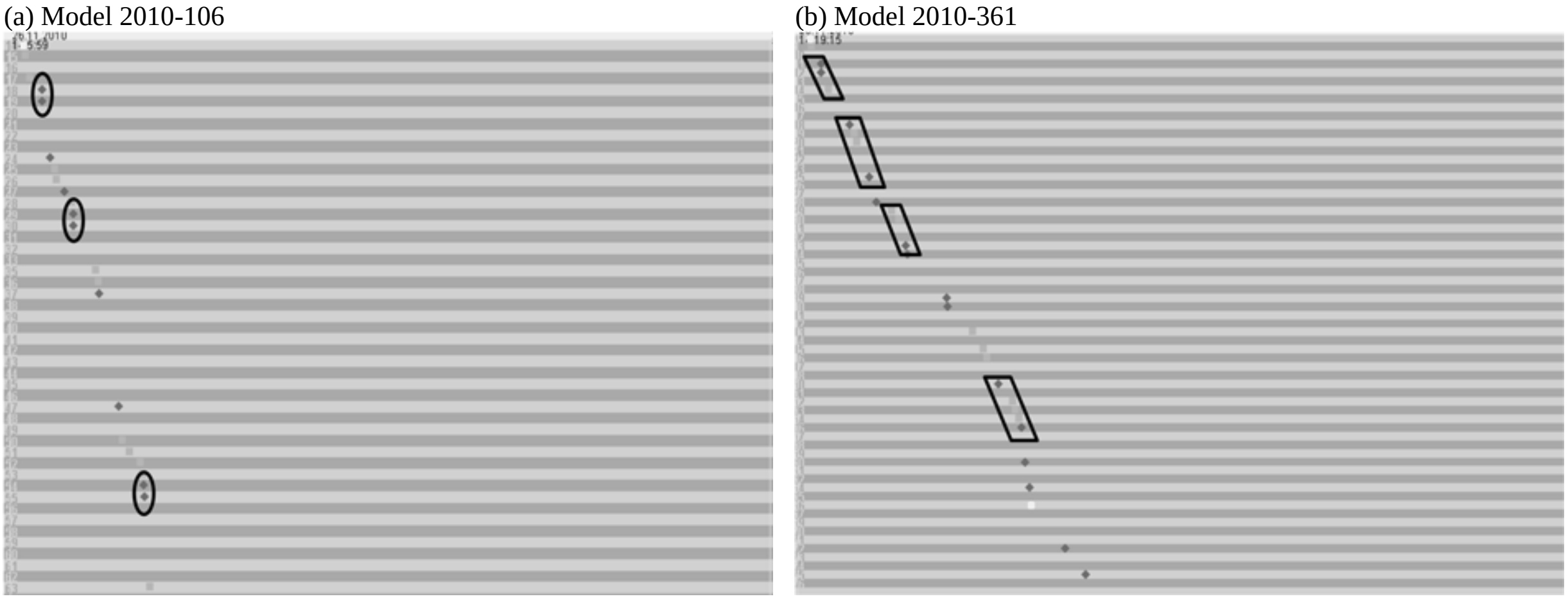

Fig. 10. Order of creation of gateways and activities.
(charts are sorted by First operation and hide edges, moves, deletes and renames)

#### 4.3.5 Patterns of create operations: chunked modeling

***Observation***: Some charts contain groups of different shades of green delimited by pauses (see Section 4.2).
***Interpretation***: Some modelers work in chunks: i.e., they alternate between creating a group of activities, gateways and edges, and pausing.

***Possible explanations***: It is observed that some modelers group the creation of parts of the model. While the second observation in Section 4.2 concludes that modelers might take a pause at various times, we observe that some modelers seem to pause only after finishing a specific part of the process model consisting of gateways, activities and edges. Possibly, these parts correspond with process model blocks (i.e., part of the model consisting of a split and matching join gateway and all intermediate nodes). Because the parts delimited by pauses seem to represent deliberate parts of the process model, we call it chunked modeling. In Fig. 11a the modeler seems to have constructed the process model in one chunk, while in Fig. 11b the modeler worked in smaller chunks.

(a) Model 2010-201 (b) Model 2010-189

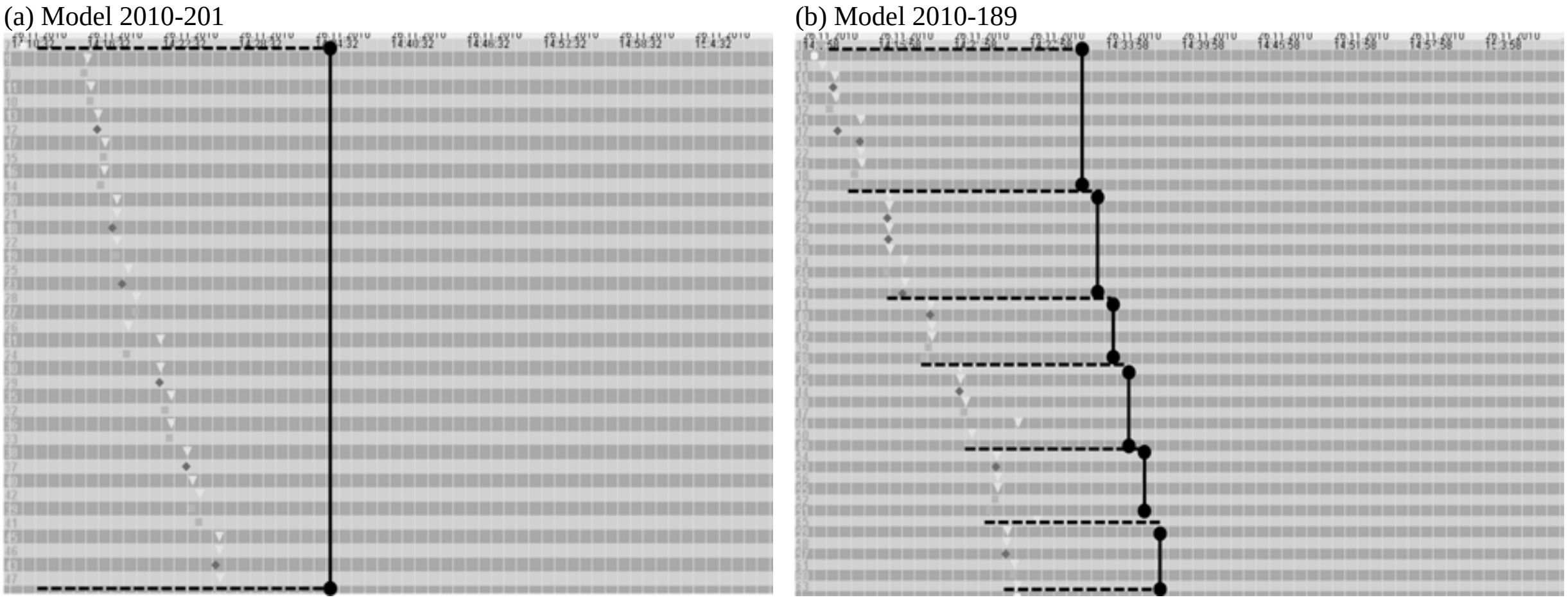

Fig. 11. Chunked modeling.
(Time intervals are set to minutes, charts hide moves, deletes and renames)

### 4.3.6 No clear patterns

***Observation***: In some charts no clear patterns are observed.
***Interpretation***: Some modelers have a more chaotic and unstructured modeling process than other modelers.

***Possible explanations***: To be complete, it is important to notice that not every modeler has a clear, observable modeling approach. Some modelers work rather chaotically and tend to work on different parts of the model simultaneously. Possibly, they are not determined about how to construct the model. Indeed, it can be observed in the dataset that they usually have a lot of move and delete operations. Alternatively, some modelers might be more chaotic in nature regardless there level of experience. This might be checked if demographic data about the *need for structure* (Thompson, Naccarato, & Parker, 1989) are collected. Fig. 12 shows examples.

(a) Model 2010-258 (b) Model 2010-270

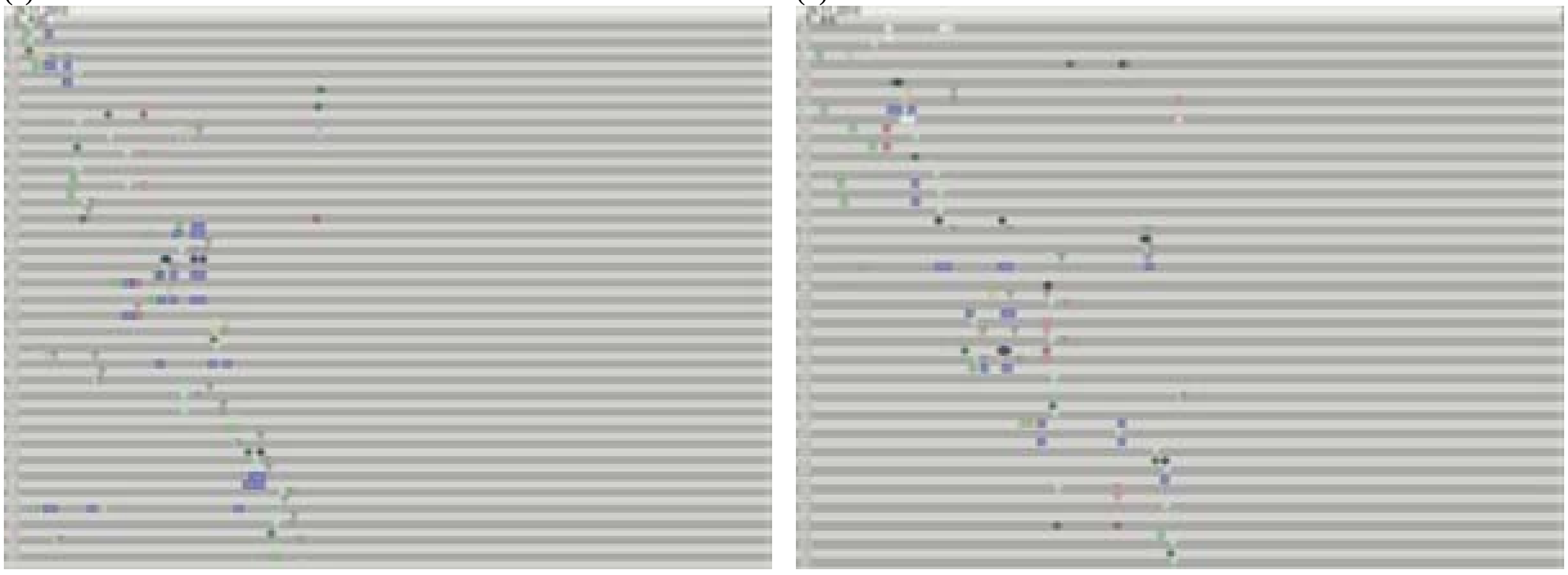

Fig. 12. Chaotic process of process modeling.
(charts are generated with default settings)

### 4.4 *Observations based on multiple diagrams*

In this section, it is demonstrated how a deeper insight can be obtained through the comparison of a PPMChart with other PPMCharts of modeling sessions for more extensive cases (see Section 4.4.1), with other PPMCharts from the same modeler (see Section 4.4.2) or with the constructed process model that corresponds to the PPMChart (see Section 4.4.3).

#### 4.4.1 Comparison between PPMCharts of different cases

***Observation***: The same patterns (as described above) occur in charts for different cases.
***Interpretation***: The same modeling approaches can be observed for different cases.

***Possible explanations***: Possibly, the discovered patterns correspond with general modeling approaches that are (rather) independent of the case to be modeled. Fig. 13 shows two examples. The left column contains examples from the pre-flight case. The right column displays examples from the more extensive mortgage case. It can be observed that similar patterns exist in both datasets. An inspection of the corresponding process models can reveal insights on potential relations between certain patterns and process model quality (see Section 4.4.3).

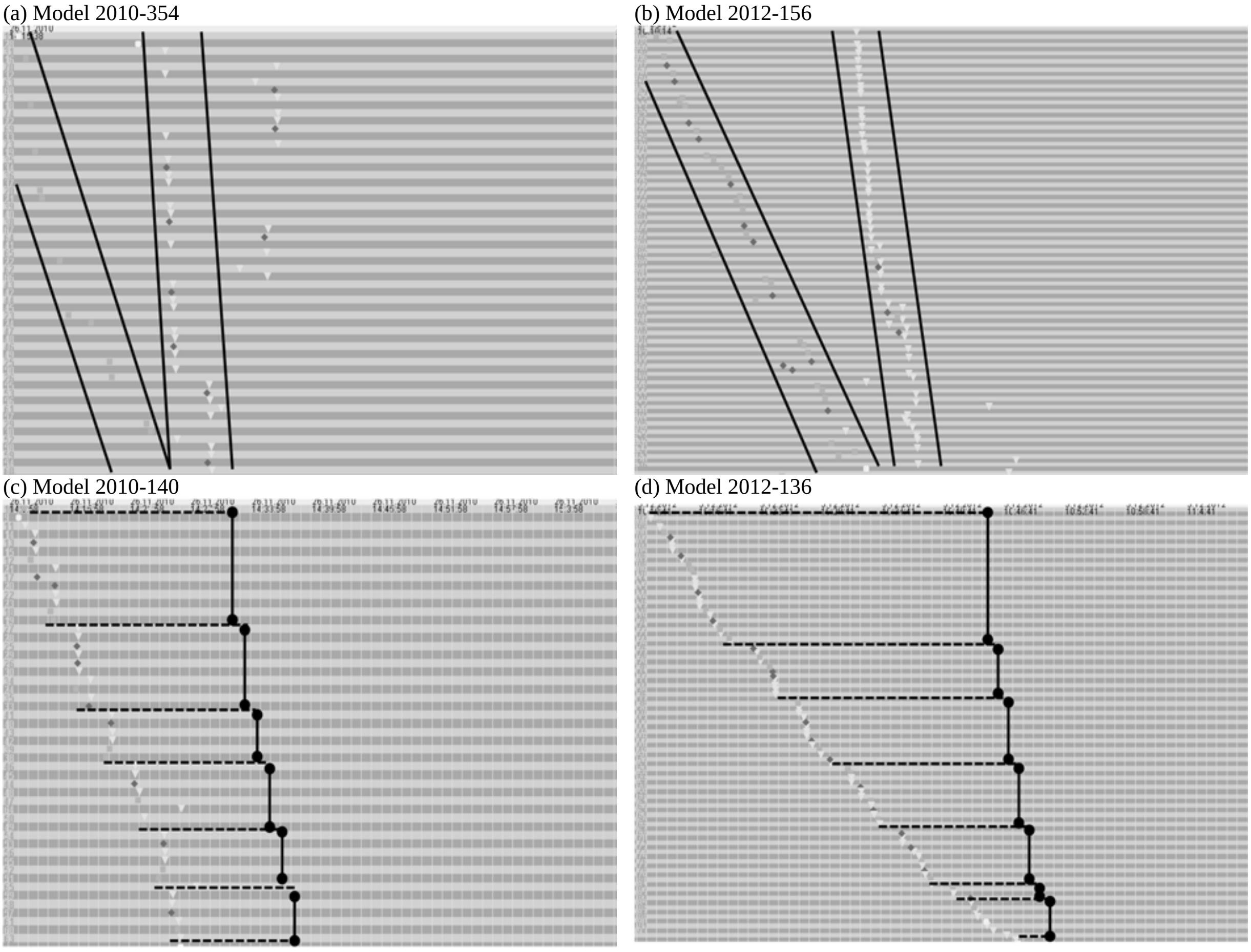

Fig. 13. Similar patterns of creation of elements in simple (a, c) and extensive cases (b, d).
(charts hide moves, deletes and renames; Time intervals of charts c-d is set to minutes)

#### 4.4.2 Comparison between PPMCharts of the same modeler modeling different cases

***Observation***: Sometimes, when the same modeler creates a process model for different cases, each chart for that modeler contains similar patterns.
***Interpretation***: Some modelers used a certain modeling approach consistently.

***Possible explanations***: Possibly, these modelers have adopted a certain individual modeling approach that can be recognized in the PPM instances for different cases. Each PPMChart at the right in Fig. 14 belongs to the same modeler as the corresponding chart at the left. Similar patterns can be observed.

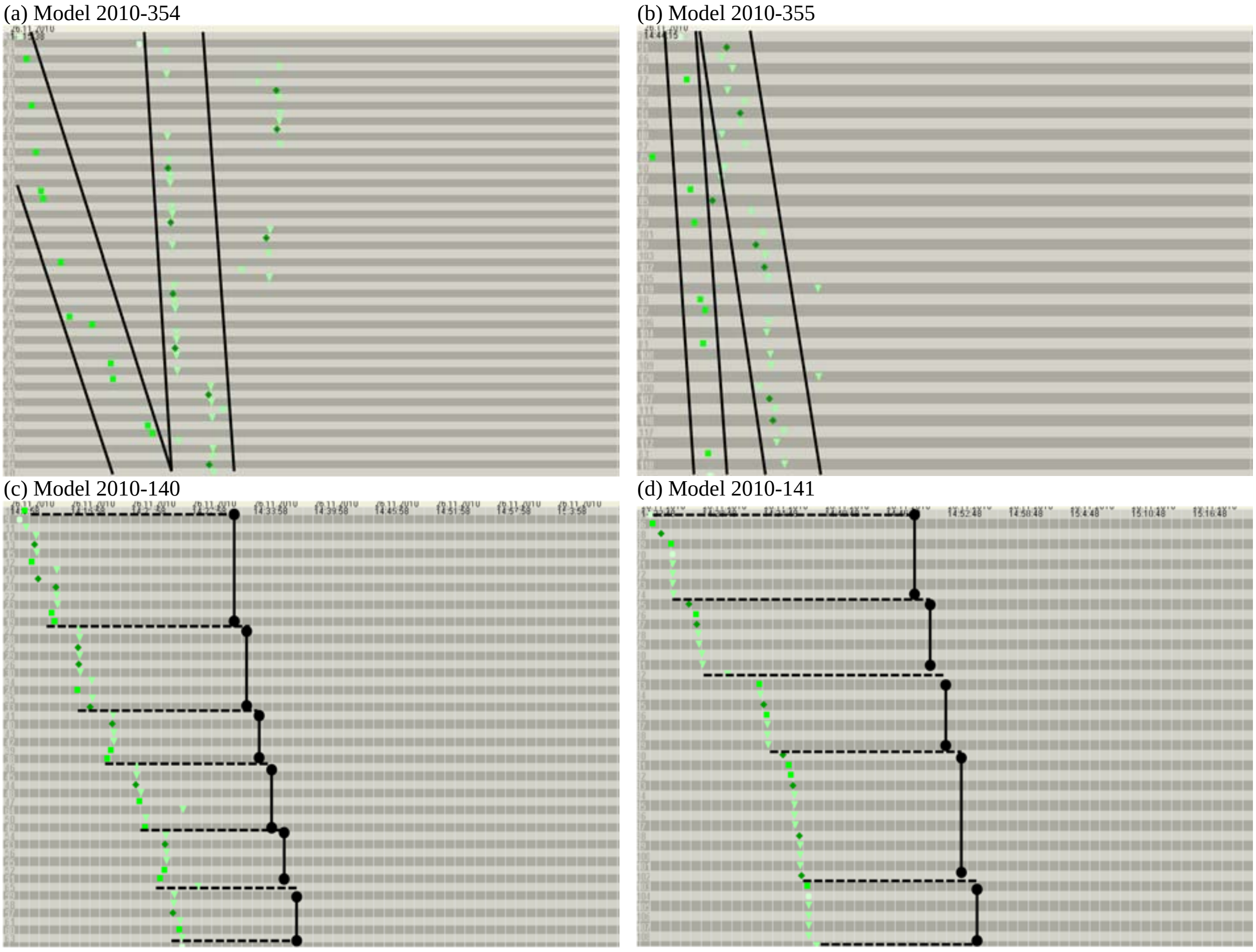

Fig. 14. Similar patterns of element creation in a first (a, c) and second (b, d) modeling effort of the same modeler. (charts hide moves, deletes and renames; Time intervals of charts c-d is set to minutes)

### 4.4.3 Comparison between PPMCharts and process models

***Observation***: Some specific patterns seem to occur only in combination with particular process models. ***Interpretation***: Modeling approaches can be discovered with the aid of PPMCharts that seem to have a relation with the properties of the modeling result (i.e., the constructed process model).

***Possible explanations***: Fig. 15 illustrates the comparison between PPMCharts and process models. The models at the right correspond to the charts at the left. The modeling approach in Fig. 15a-b is aspect-oriented (i.e., first creating events and activities, only afterwards gateways and edges). The PPM instance depicted in Fig. 15c-d shows chunked modeling. These are two different examples of a modeler that uses a clear modeling strategy that the modeler consistently applied. Note, how the corresponding process models also have a clear structure[14]. Because this particular relation can be observed in a high number of examples, this might indicate that having a clear modeling style (i.e., the patterns as described above are clearly recognizable), has an impact on the structuredness of the resulting process model.

[14] We are aware of the fact the models are unreadable. This does not prevent to judge the structure of the models. The process models can be downloaded in high resolution from http://www.janclaes.info/papers/PPMISeB.

(a) Model 2010-354 (b) Model 20120-354

(c) Model 2010-140 (d) Model 2010-140

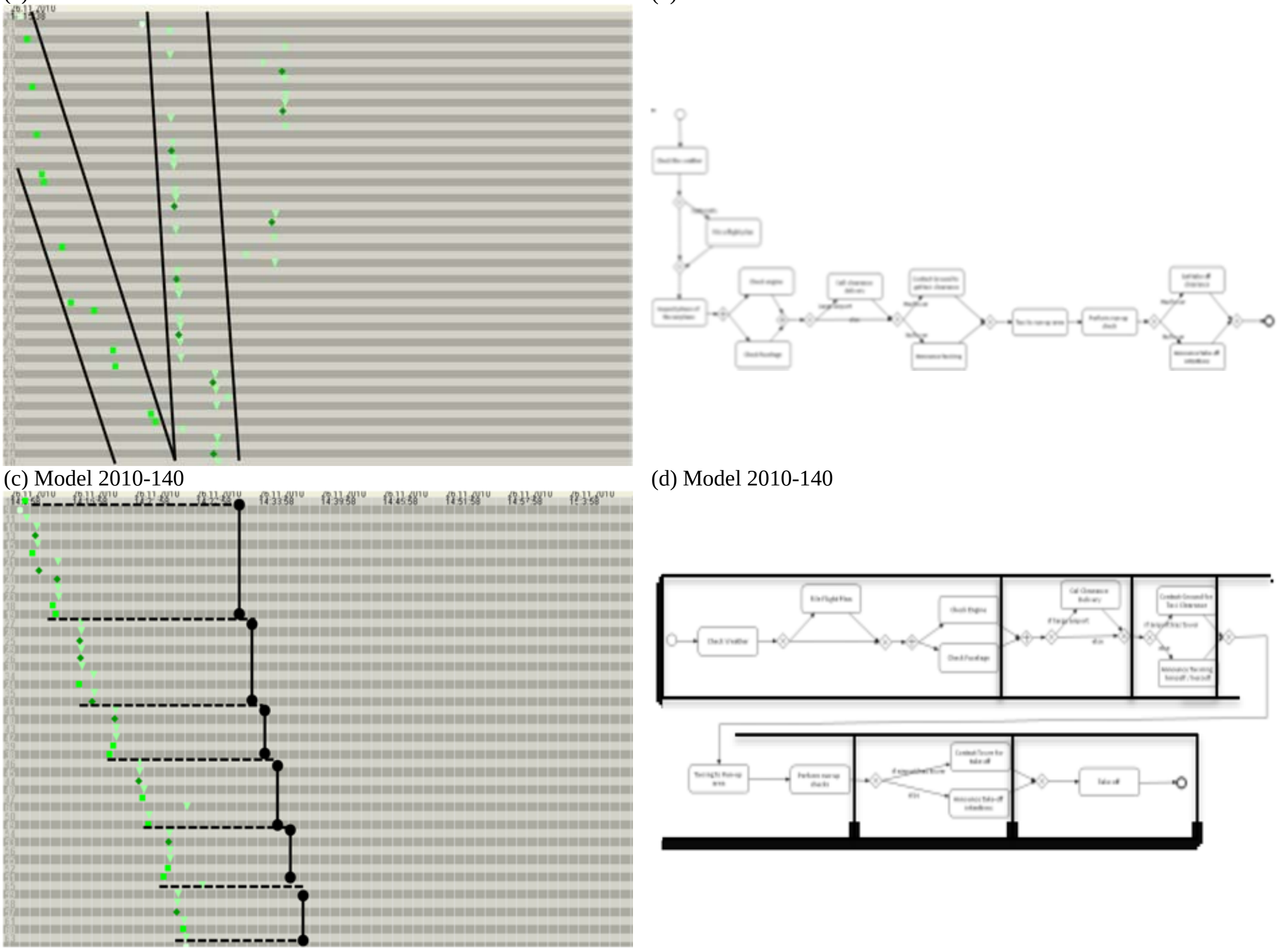

Fig. 15. Patterns of creation of elements (a, c) and corresponding process models (b, d).
(charts hide moves, deletes and renames; Time intervals of chart c is set to minutes)

# 5 Qualitative evaluation

This section reports on a qualitative study that confirms the usefulness of the PPMChart and the increase in cognitive effectiveness compared to the Dotted Chart. The evaluation was designed to collect mainly qualitative data about the use and usefulness of the visualization. Participants to the evaluation (see Section 5.1) were instructed to use a particular visualization (see Section 5.2) to graphically represent data about the process of process modeling (PPM) (see Section 5.3) and study the data for one hour (see Section 5.4). During the exercise qualitative and quantitative data were collected and before and afterwards a number of questions were posed to collect additional data (see Section 5.5). The results of the empirical evaluation are presented in Section 5.6.

### *5.1 Participants*

The PPMChart visualization is in the first place intended to support explorative research into the PPM. The pool of potential participants therefore consisted of researchers that were familiar with process modeling, but that were not involved yet in research about the PPM. Six academic researchers with varying levels of experience volunteered to participate (three PhD students and one scientific programmer from Ghent University; and one PhD student and one assistant professor from Eindhoven University of Technology).

### *5.2 Visualization tools*

Two visualizations were compared. The Dotted Chart in ProM 6.2 (see Section 2.1) was compared to the PPMChart (see Section 3.2). Each participant had to work with only one visualization (i.e., either Dotted Chart or PPMChart). To prevent interference, one user that was familiar with the Dotted Chart was instructed to use the PPMChart. A random visualization was appointed to the other users, but in such a way that each visualization in the end would have been tested by an equal amount of users.

### 5.3 *Input data*

The evaluation involved five datasets that originate from a single observational modeling session. Every participant got access to all five datasets. These contained the information of the construction of models 2010-140 (see Fig. 13c), 2010-213 (see Fig. 7b), 2010-270 (see Fig. 12b), 2010-354 (see Fig. 13a), and 2010-361 (see Fig. 10b). They were selected as outlier examples, which was explicitly mentioned to the participants. Each dataset consisted of a pre-loaded event log in ProM according to the format described in Section 3.4.1 and the corresponding process model on paper. It was explained to the participants that the data originated from a single modeling session where five different students had to model the same case using the same case description. No further information was provided about the modelers, the case, the quality of the process model, etc.

### 5.4 *Protocol*

The PPMChart visualization was developed to support explorative research. Therefore, no particular task was prepared, but the participants were simply instructed to use the visualization and the five event logs for one hour and to try to discover as much information as possible about how people construct models. Also, because of the explorative nature of the intended use, we selected researchers that had no experience with the visualization and we provided only minimal tool training.

The participants were asked to take place behind a computer on which ProM 6 and screen recorder software were pre-installed and running. The screen recording was activated as soon as they agreed with the capturing of data about their usage of the computer. Voice recording was activated accordingly after reception of their permission. First, it was briefly explained that the data was collected from observational modeling sessions with students and without treatment. The level of detail of the recorded data was described using some examples (e.g., creation of an activity, creation of an edge, movement of the activity, etc.). Next, ProM was introduced briefly. The five event logs were already imported in ProM. Using one random event log from the series of five, the assigned visualization was introduced. It was shown with the example event log how to generate the chart for a selected event log in ProM and which configuration options existed in the visualization plug-in.

Before starting the exercise, users were asked to think about the focus of their analysis. Using only the information of the brief introduction of the data and tool, they were instructed to elaborate on what kind of information they expected to be able to extract from the visualization. The introduction, instruction and focus question took no more than half an hour for each participant.

During the assignment, participants got one hour to explore the tool and the data and to reveal as much useful information about "how do people model" as possible. The initial exploration of the tool by the participant was included in the one hour duration, because the influence of the intuitiveness of the tool on the results was desired to be reflected in the results. However, to simulate real use conditions participants were allowed to ask the administrator for help if they did not understand a feature or did not know how to set a particular configuration. While they were working, participants were asked to think out loud and to clearly describe relevant observations. If they described an observation uncarefully, the administrator asked for a more precise description. Approximately every 15 minutes, when the administrator felt it would interrupt the participants the least, the session was paused for some minutes to ask two questions: (i) what is the most relevant observation so far, and (ii) from all the possible information that is present in the data and could be discovered with the visualization, how much percent do you think you already discovered. These questions served as a short mental break and were prearranged to keep the participant focused on the goal to derive as much insight as possible.

After exactly one hour the session was closed and the participants were asked to comment on the visualization. They were explicitly asked to be critical, to think of what bothered them, and to suggest improvements. When the participants could not think of any more feedback, the evaluation session was concluded with a short debriefing. They were explained that they used the developed or existing visualization and some information was given about our own initial insights into the PPM as presented in Section 4.

### 5.5 *Measurements*

Qualitative and quantitative data were collected during the evaluation session. Qualitative data includes the reported observations of the participants, observations of the administrator in how the participant uses the visualization, opinions of the participant about the data, visualization and tool (expressed during or after the assignment). Quantitative data includes per observation time and domain value (score from 1 to 5) assessed by an external BPM(N) expert. It was also coded by the authors whether an observation was correct, was directed or unexpected (based on the focus question answers prior to the assignment), was either focused on depth or breadth.

### *5.6 Results*

To evaluate whether the PPMChart visualization is useful and cognitively more effective than the Dotted Chart, qualitative data was collected through participant observation and interviews during the evaluation session (Myers, 1997).

#### 5.6.1 Usefulness

Section 4 clearly demonstrates the usefulness of the visualization for our own previous research. The question that remains is whether other users could draw meaningful conclusions from the charts as well.

It was observed that all three researchers that used the PPMChart visualization in the assignment indeed made similar observations than the ones described in this paper. For example, they reported on patterns of deletion, movement and creation and discovered the styles that we labeled flow-oriented and aspect-oriented modeling. One of the participants even described a third similar style, which we did not discover in that dataset yet: one modeler appears to first have modeled one path from start to end and later on added all the exceptional paths using XOR gateways.

During the exercise and from the interview with the participants, perceptions about the tool were captured. The three researchers that tested the PPMChart visualization were mainly positive about the tool ("*very complete visualization*", "*contains a lot of useful options*", "*handy overview*", "*intuitive colors and shapes*"). On the question to name suggestions, participants proposed to "*display the model number somewhere on the screen*", "*reformulate filter options from 'hide' element to 'show' element*". Two of the participants mentioned that "*it was difficult in the beginning to use the tool because of its extensive options*". On the other hand, they both felt confident using the tool after less than an hour (i.e., before the end of the exercise).

Summarized, with the aid of the PPMChart visualization all the patterns and conclusions that are presented in Section 4 were discovered by the participants within one hour (using the five purposefully selected event logs). Also, some additional insights were derived from the charts (e.g., happy path first modeling style). The users perceive (parts of) the tool as being "*complete*", "*useful*", "*handy*" and "*intuitive*".

#### 5.6.2 Increased cognitive effectiveness

We acknowledge that the Dotted Chart can be used for the study of the PPM, but we claim the PPMChart supports such research in a more effective way from a cognitive viewpoint. To evaluate this, the observations and feedback of the participants using both visualizations are compared.

First, it should be noted that we did not evaluate the choice of representing only one PPM instance at the time. Both participants appointed to the Dotted Chart and appointed to the PPMChart were provided with the same event logs containing one PPM instance each. This non-typical use of the Dotted Chart does however not cause a substantial bias, because none of the participants were familiar with the visualization they used. Therefore, they did not know that Dotted Chart is typically used with an event log containing multiple process instances.

All three researchers working with the Dotted Chart gradually started to change colors in the chart. Colors were picked to focus on a specific operation (i.e., because of a lack of semantic transparency), to clearly distinguish between different operations (i.e., to increase perceptual discriminability), or to give similar operations a similar color using different shades of the same color (i.e., to improve visual expressiveness, perceptual discriminability and graphic economy). They used an individual fixed color scheme in the end and started their last analysis by first changing the colors to their specific color scheme. In contrast, in the use of the PPMChart no colors were changed at all, which indicates the fixed color scheme of the PPMChart is intuitive, distinguishable and expressive (enough).

It was also observed that every Dotted Chart that was generated had an initial zoom level that did not fit the needs of the user (i.e., they changed it immediately after changing the colors). In PPMCharts zoom operations were only detected in the course of analysis to focus on a difficult zone.

In a PPMChart the timelines are sorted by the logical execution order of the process model from start event towards end event (i.e., by Distance from start). This makes it easier to connect the information to the process model on paper (i.e., cognitive integration). Only rarely the sort order of the PPMChart was changed by the users. In contrast, we recorded substantially more changes in the sort order for the Dotted Charts (i.e., to increase cognitive fit). This suggests the default sort order from the PPMChart can be considered to add value for the user.

Finally, it was noted that participants to the evaluation (i.e., using the Dotted Chart) occasionally changed the color of one particular type of event in a strikingly different color. We deduce that those participants wanted to focus only on this type of event, while disregarding the other events (i.e., to increase cognitive fit and for complexity management). In the PPMChart we did not observe this behavior, but two of the three participants used the filters in the plug-in for similar analysis tasks.

When we asked the participants about the Dotted Chart, the main suggestions pointed to "resolve *weird zooming*", "*colors are different from previously generated chart*", "*hard to link back to the process model*", "*metric panel is useless*", "*drop the useless options*", "*use tones of colors*", "*zoom to fill the initial screen*", "*I can't tell which events are on the same model element*". Other suggestions were to "*work with predefined sets of configuration options*", "*use multi gesture zooming*".

No substantial difference in the amount or quality of derived insights between the two visualizations was observed. But in the Dotted Chart users put a lot of effort in first configuring the tool to its maximal cognitive effectiveness to support their analysis, which we did not observe in the PPMChart. Participants reported on several cognitive drawbacks of Dotted Chart concerning for example cognitive integration ("*different colors*", "*hard to link*") and semantic transparency ("*which events are on the same model element*").

# 6 Limitations, implications and future research

### *6.1 Limitations of this study*

The usefulness of the visualization is illustrated in Section 4. The presented examples mostly represent extremes. An informal analysis indicated that, although these extremes were present in the dataset, for a number of observations there appears to be a continuum of examples in between the presented extremes. Furthermore, these examples are based on data of modeling sessions with students only. They illustrate the usefulness of the visualization, but must not be considered to be representative for all modelers. Therefore, a profound study is needed to examine the circumstances and the generalizability of the observations and the possible explanations, which is out of the scope for this paper. Nonetheless, these examples provide a useful understanding of how the visualization can be applied for discovering properties of the process of process modeling (PPM).

The evaluation of the improved cognitive effectiveness of the PPMChart against Dotted Chart includes the study of qualitative data collected in an empirical evaluation study. Case study research could be performed to examine more in-depth and in a more realistic setting how the tool supports explorative PPM research in a cognitive effective way. Nevertheless, a summary of exemplar cases that show how the PPMChart could help or has helped researchers study the PPM is provided in Section 6.2 to exemplify the implications for research. In order to evaluate the difference in speed, amount and quality of derived insights between both visualizations, a quantitative approach would be desired. However, for a reliable quantitative evaluation a higher number of participants are required, which is cumbersome given the limited number of people in the intended target group. Moreover, the level of understanding of the reported insights from the point of view of the participant is largely lost when quantified (Kaplan & Maxwell, 2005), which makes a quantitative study less suitable to evaluate a tool for explorative research.

### *6.2 Implications for research and practice*

The comparison of different diagrams and the discussion section clearly show that the PPMChart visualization can be a very helpful instrument in the exploration of data from observational modeling sessions. Concrete examples in Section 4.4 illustrate that the discovered patterns might relate to modeling approaches that are general (independent of the case), individual (dependent of the modeler) and that have an impact on the properties of the resulting process model (see Section 6.2.1). If this interpretation is correct, this means that the PPMChart can facilitate the study of the PPM significantly. In particular, this would be very useful for the research into modeling approaches that have a positive impact on process model quality (see Section 6.2.2). This way, it facilitates also the improvement of training or tool support to increase model quality (see Section 6.3). Due to the growing importance of process models in process analyses and optimization efforts, this is an important contribution to the research and practice of the business process management field.

#### 6.2.1 Discovery of modeling phases and styles

When modelers work aspect-oriented, the PPM proceeds in phases. The PPMCharts for our dataset revealed that some PPM instances contain, besides the *creation phases*, also *deletion phases* (see Section 4.3.1), or *move phases* (see Section 4.3.2). Sometimes, the phases are interspersed with pauses, which can be easily detected graphically in the PPMCharts (see Section 4.2). Furthermore, the study of PPMCharts proposes that the *modeling phases* in one PPM instance might be aspect-oriented (i.e., during a modeling phase the modeler concentrates subsequently on different aspects: e.g., first content, then structure; see Section 4.3.3) or flow-oriented (i.e., the different modeling phases correspond with the creation of model chunks that are finished one after each other; see Section 4.3.4).
The existence of phases in the PPM was discovered before with the aid of Modeling Phase Diagrams (see Section 7.1) representing data from a different case than used for the datasets described in this paper (Pinggera, Zugal, et al., 2012). Three phases are distinguished: a modeling phase in which the modeler mainly creates new elements, a comprehension phase in which the modeler pauses his modeling activities, and a reconciliation phase in which the modeler moves and deletes elements. Pinggera, et al. report on a study that analyses the modeler's eye movement in the model canvas and case description areas on screen (Pinggera, Furtner, et al., 2013). It was concluded that some modelers use the pause to reflect on their model so far and some modelers use the pause for reading the case description.

Next, similar patterns were observed between the PPM instances of the same modeler for different cases (see Section 4.4.1) and between instances of different modelers and different cases (see Section 4.4.2). This suggests that common modeling styles exist, and that a modeler might adopt a rather fixed modeling style over time.
Pinggera, et al. used clustering techniques on the dataset from the modeling session in Eindhoven in 2010 and discovered three modeling styles: "*(i) modeling with high efficiency, (ii) modeling emphasizing a good layout of the model, being created less efficiently, and (iii) modeling that is neither very efficient nor very focused on layouting*" (Pinggera, Soffer, et al., 2013). The three styles can be characterized by the modeling speed (i.e., fast versus slow) and the amount of reconciliation operations (i.e., many versus few moves, deletes, renames).

This clearly demonstrates the usefulness of the PPMChart visualization. It can be used to make the same observations that have lead to the interesting insights presented in this section (i.e., the existence of phases and styles). Furthermore, concerning patterns of individual operations within phases, it provides more detailed information than for example Modeling Phase Diagrams.

#### 6.2.2 Link with process model quality

While the previous section discusses how the PPMChart could have been used in other research about the PPM, this section describes how the PPMChart was actually used in PPM research. Because of the recent development of the PPMChart, only a limited part of our findings is already published in academic articles. The comparison of PPMCharts with the corresponding process models (see Section 4.4.3) revealed a potential link between certain PPM properties and process model quality. For example, we observed that modelers with a short PPM (see Section 4.2) tend to produce process models of better quality. It was also shown that a lot of move operations (see Section 4.2) often goes hand in hand with models of poorer quality. Finally, we discovered that modelers with a chunked modeling style (see Section 4.3.4) typically create better process models. These three conjectures were further studied and a statistical analysis of the data confirmed that these relations exist in the dataset (Claes et al., 2012).

### *6.3 Future research*

To be able to reach the point of gaining knowledge of the connection between the characteristics of the PPM and the quality of the resulting process model, a substantial amount of research still needs to be performed. Concerning the graphical representation of the collected data in the PPMCharts presented in this paper, the future work concentrates on improving the support for the inclusion of more data into the analysis (i.e., improve the cognitive integration of information). For example, from our experience in using the charts for research, we learned it would be valuable to simplify the way to link individual dots in charts with elements in the corresponding process models. Additionally, the generation and study of multiple PPMCharts at once can be optimized. For example, it is now not possible to produce multiple PPMCharts (i.e., for multiple PPM instances) at once or to configure multiple charts at once. Tools may also help in comparing different charts in a visual way. Research is needed to examine how this can be optimally implemented.

Furthermore, it might be interesting to include information about the *modeler* (e.g., is there a difference between the PPM for novice and expert modelers), *model language* (e.g., do certain modeling notations influence the way in which a modeler constructs the model), *modeling tool* (e.g., how do existing tool features help the modeler), *modeling case* (e.g., how much does the complexity of the case to be modeled influences the PPM), etc. in the charts in such a way that the additional information does not lower the cognitive advantages of the visualization. Future work may also include the extension of the configuration options, such as representing other operations than mentioned in Table 1, supplementary filter options, sort options, etc.

# 7 Related work

This section describes related work from several perspectives. First of all, related research in the area of visualizing the process of process modeling (PPM) is summarized (see Section 7.1). Next, the usage of traditional process modeling notations for the visualization of the PPM is discussed (see Section 7.2). Subsequently, a selection of other visualizations that focus on hierarchy and control perspectives is presented (see Section 7.3).

## *7.1 Visualizations of the process of process modeling*

Pinggera, et al. (2012) use Modeling Phase Diagram visualizations to graphically represent the consecutive phases of a modeler in the construction of a process model. A Modeling Phase Diagram is a line chart representing the number of elements present in the partial process model during the model construction. The alternating line color (i.e., black or grey) and line style (i.e., solid or dotted line) indicate the consecutive phases of the modeling process (see Fig. 16a). While this visualization is very useful to provide a consolidated graphical overview of these modeling phases, it does not allow zooming in on the singular recorded events of the PPM instance it represents. It is tailored to the study of consecutive phases in the PPM. Alternatively, the PPMChart visualization (see Fig. 16b) was constructed to represent the captured data at a detailed level, which facilitates the analysis of singular operations performed by the modeler, rather than analyzing the characteristics of modeling phases. The result is a more comprehensive visualization than the Modeling Phase Diagrams. Moreover, the PPMChart displays the data as it was recorded and leaves the analysis and recognition of patterns, and the interpretation about the cause and meaning of specific operations to the reader of the chart. While the Modeling Phase Diagram is developed to support a very specific analysis at an aggregated level (i.e., of the model phases in the PPM), the PPMChart is designed to support the analysis of various aspects of the PPM at a detailed level.

(a) Modeling Phase Diagram (b) PPMChart

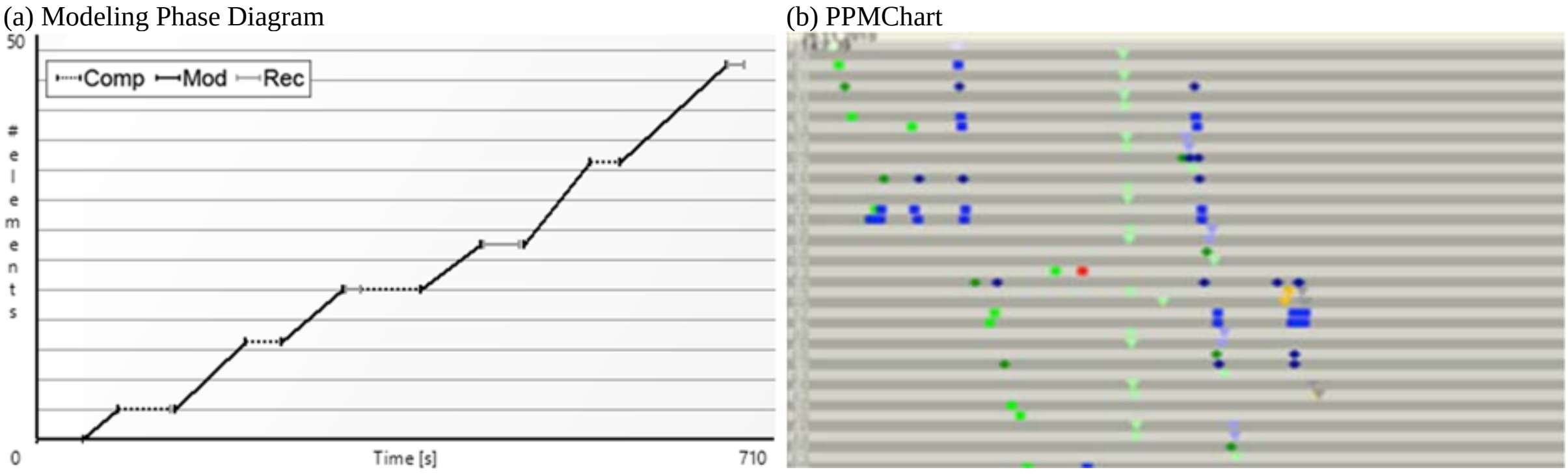

Fig. 16. Example of a Modeling Phase Diagram and PPMChart for the same PPM instance.

## *7.2 Process models*

A conceptual (process) model is a "*formal description of some aspects of the physical and social world around us for purposes of understanding and communication*" (Mylopoulos, 1992, p. 3). A combination of textual and graphical notations can be used for this formal specification (Engelen & Van den Brand, 2010). When the focus is on control flow (i.e., the order of activities and events), graphical representations are preferred over textual models (Weske, 2007). Various graphical process model notations exist (e.g., BPMN (OMG, 2011a), UML Activity diagrams (OMG, 2011b), Petri nets (Reisig & Rozenberg, 1998), Workflow nets (Van der Aalst, 1998), YAWL (Van der Aalst & Ter Hofstede, 2005), and Event-Driven Process Chains (Scheer, 1998)). To capture the complexity of processes, process models make use of the principles of structuring and abstraction (Polyvyanyy, 2012). For example, many process models do not show information at instance level, but summarize information of different executions, while often hiding infrequent behavior (Polyvyanyy, Smirnov, & Weske, 2010; Reichert, 2013).

In terms of process model visualization, it should be noted that recent research studies how changes in the process model notation can help to lower the mental effort for the reader to interpret the models. For example, in order to bridge the gap between the traditional formal process model notations used by process modelers and the informal representations often used in practice (Barros & Ter Hofstede, 1998; Phalp, 1998), icons are introduced in existing model notations (Mendling, Recker, & Reijers, 2010). Other research focuses on the addition of a third dimension to the model (Effinger, 2013) or represents the process model in a virtual world (Brown, 2010; Guo, Brown, & Rasmussen, 2013). Also, more creative approaches are used, such as sonification (Hildebrandt, Kriglstein, & Rinderle-Ma, 2012).

Although the different abstraction mechanisms and visual optimizations of the notations have success in supporting readers of the models to deal with the complexity of the represented process, they have in common that they hide details of individual instances. Therefore, classical process model notations are not suitable for the in-depth analysis at instance level of processes in general and the PPM in particular. In contrast, the PPMChart visualization represents only one PPM instance of which no details are hidden. This only shifts the complexity problem, because one still needs to compare a high number of charts for obtaining an overview over different instances, but this better fits with the goal of the exploratory research of this paper, which aims at revealing in-depth information. Therefore, traditional process model notations are not appropriate to support this research.

### *7.3 Visualizations that concentrate on control flow and hierarchy*

An optimal visualization technique for exploratory research about the characteristics of the PPM should show a lot of details about the recorded operations on model elements, focusing on timing and relative order and taking the hierarchical structure of the data into account. Therefore, this subsection presents a number of visualizations that concentrate on control flow or hierarchy and discusses their potential usefulness as an instrument to graphically detect characteristics of the PPM from the collected observational data.

Different visualizations exist to represent *hierarchical* information (e.g., Treemaps (Johnson & Shneiderman, 1991), Timeline Trees (Burch, Beck, & Diehl, 2008), Arctrees (Neumann, 2005), Information Slices (Andrews, 1998), Sunburst diagrams (Stasko & Zhang, 2000)). These graphical representations display the details in such an hierarchical placement that relations between data elements can graphically be discovered.

Treemaps display information about elements as rectangular blocks of which position, width, height and color are the main properties to represent characteristics of the data (see Fig. 17a). They make optimal use of the available space in the chart, because the whole chart is filled with information. However, the TreeMap visualization cannot optimally support the research into the PPM, because it focuses mostly on hierarchy and relative importance of the represented data elements. It has not been optimized to provide cognitive support for the recognition of patterns in the ordering of operations. Also Arctrees, Information Slices and Sunburst diagrams, which make use of a radial placement of information visualization elements, have the same shortcomings.

Timeline Trees do include an explicit representation of order and timing of data elements grouped in categories. The hierarchy of the data categories is represented by a textual tree, and the timing of transactions for each of the data elements in a category is visualized by a timeline for each of the leaves of the tree (see Fig. 17b). The focus of the visualization is evenly spread over the hierarchy and the timing of data elements. For the PPM, this visualization can be used; although, in our opinion, the tree representation of the hierarchical structure of the data takes too much room and is distracting if the number of data categories increases.

Other well-known visualizations focus less on the representation of hierarchy, but more on the *timing* of or the *relation* between the data elements (e.g., Gantt charts (Gantt, 1913) and Railroad line diagrams (Tufte, 1983)). Gantt charts are used in project planning to analyze phases, dependencies and timing of projects (Wilson, 2003). The different phases of the project are mentioned beneath each other. Besides each phase, a horizontal bar indicates the planned timeframe for each phase on the time axis (see Fig. 17c). The focus on the length of phases and dependencies between phases, and the lack of attention to optimally represented details of individual steps of each phase, makes them less suitable for the analysis of the PPM with regard to the discovery of useful insights at the level of patterns of individual operations of the modeler.

Next, different informal Railroad line diagram visualizations exist to represent the routes and hour schedules of trains. They can be considered variants of, for example, Marey’s train schedule. The route of a train from one station to another is represented by a line that traverses the chart from left to right. Vertically the different stations are displayed and the horizontal axis represents a timeline (see Fig. 17d). This visualization is suitable for the display of information about an object of which the properties change in two dimensions (i.e., place and time), but does not allow for adding more dimensions of information easily without decreasing its cognitive effectiveness substantially, which would have been necessary to be used for the analysis of the PPM.

(a) TreeMap (from Johnson, 1993)

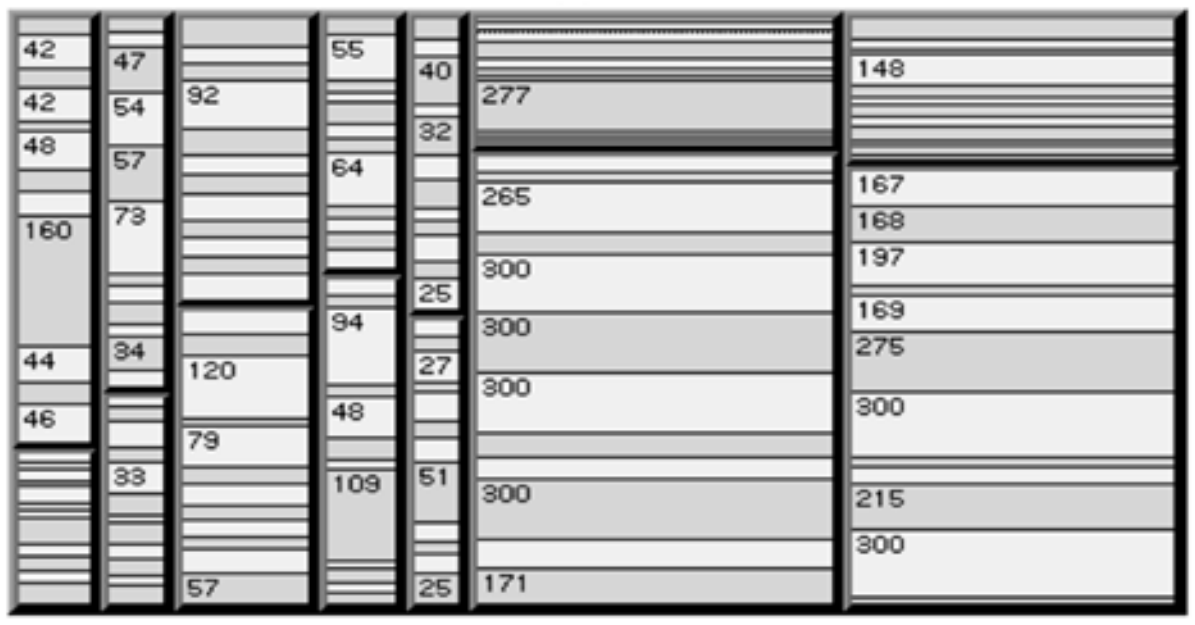

(b) Timeline tree (from Burch et al., 2008)

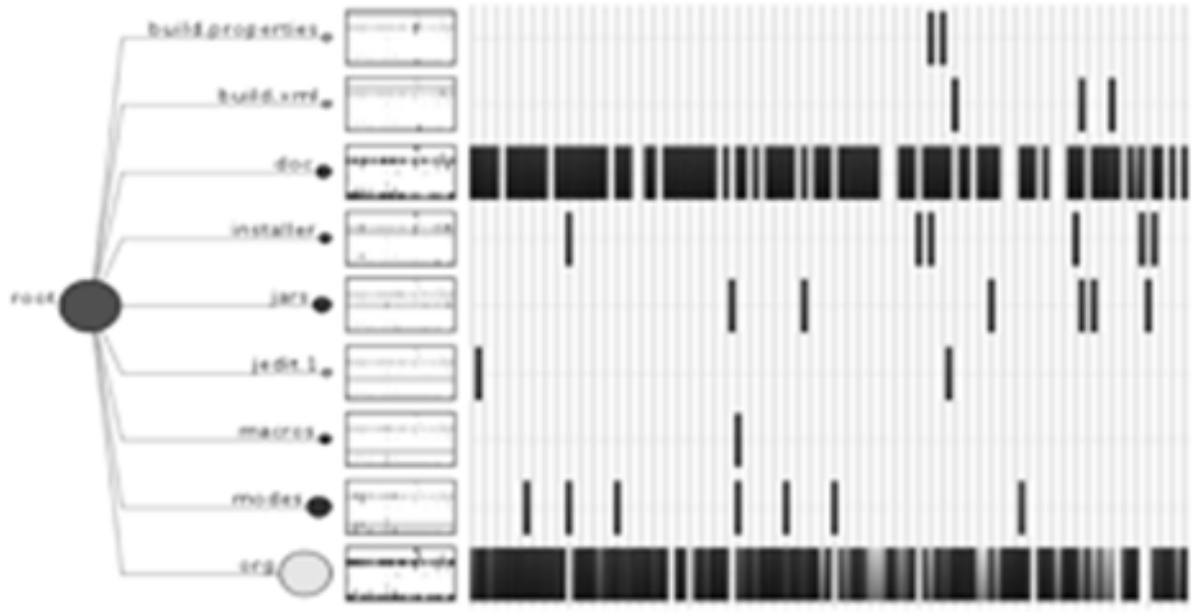

(c) Gantt chart (from SE blog[15])

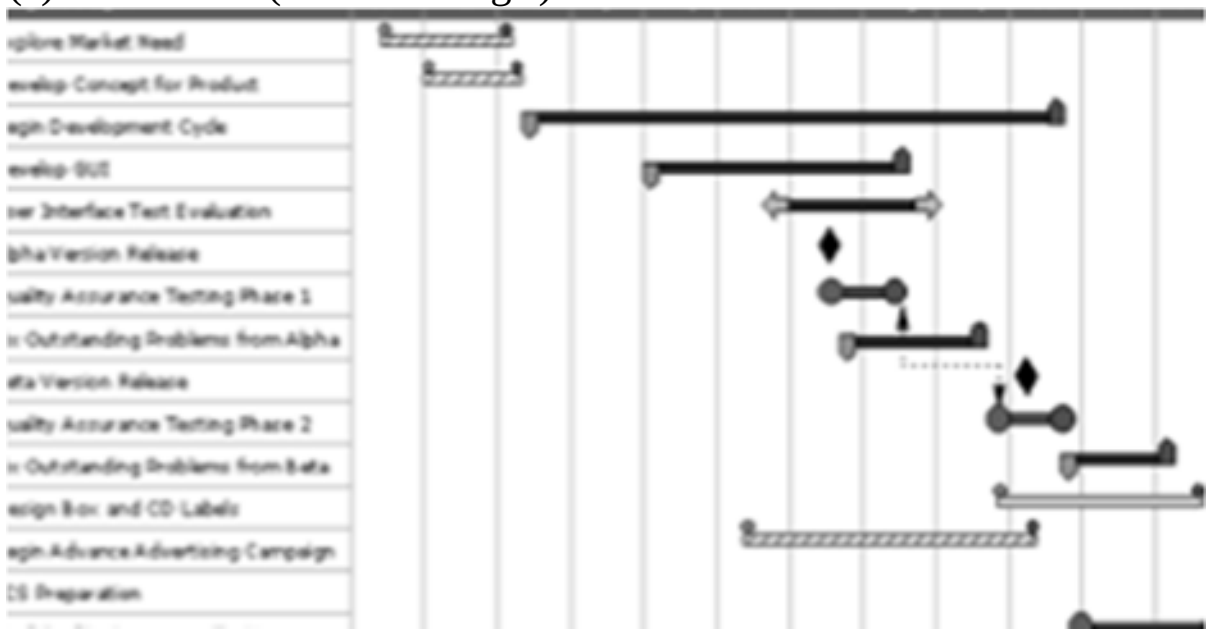

(d) E.J. Marey's train schedule (from Tufte, 1983)

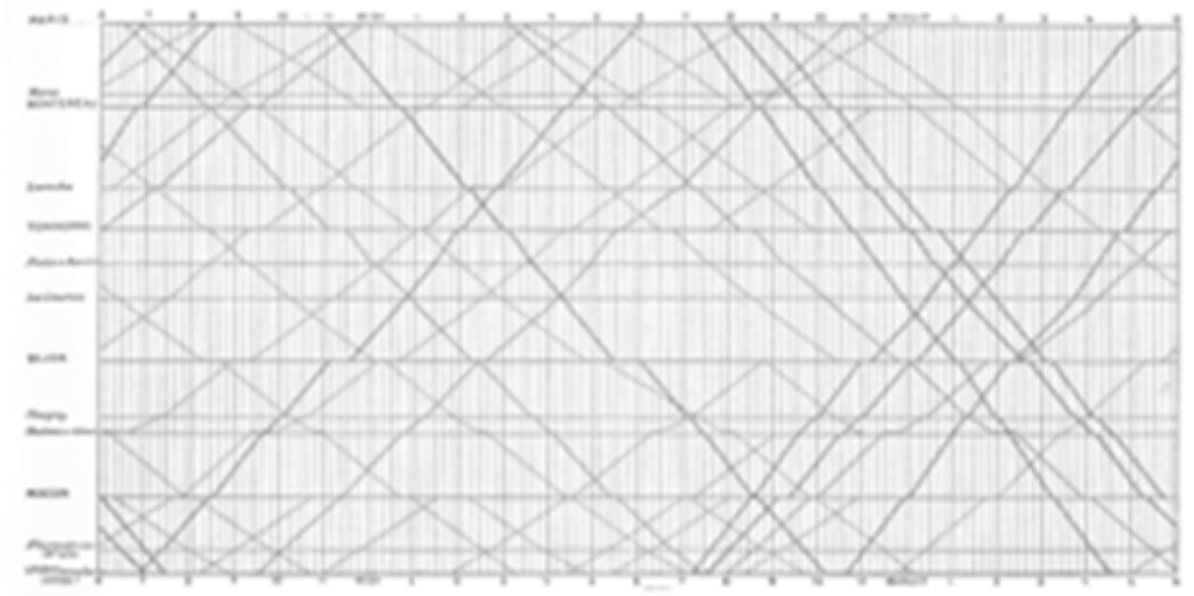

Fig. 17. Examples of process visualizations that are not considered to be traditional process models.

# 8 Conclusion

The goal of the research described in this paper was to design and implement a visualization that helps to study observational data about the process of process modeling (PPM) in a cognitive effective way. The visualization makes the characteristics of PPM instances explicit, which facilitates the development of theory, training and tool support for various aspects of the PPM, and especially in the context of increasing the quality of the resulting process models.

The PPMChart visualization of the PPM presented in this paper displays modeling operations of one modeler in the construction of a single process model as colored and shaped dots in a chart. The dots are positioned on horizontal timelines that represent the model elements on which the operations are performed. The PPMChart is implemented in the process mining tool ProM in such a way that various options can be configured and that the data can be filtered from within the plug-in. This allows to effortlessly take different views at different levels of abstraction on the modeling operations. The paper contains an extensive list of examples and observations to demonstrate the usefulness of this graphical representation in analyzing the PPM and the modeling behavior and styles of different modelers. A qualitative study confirms the usefulness of the PPMChart and the improved cognitive effectiveness compared to the Dotted Chart.

Two specific characteristics of the visualization need special attention. Firstly, the PPMCharts show raw, uninterpreted data. Each dot represents a clearly observable distinct modeling operation of the modeler. The interpretation of the meaning or cause of a specific operation is left up to the reader of the charts. Secondly, the visualization makes advantage of the same benefits that were originally present in the Dotted Chart Analysis plug-in. The presentation of operations makes use of dots that have a particular color, shade, shape and position, which means that all available data are presented in the visualization and that the reader can zoom in on details or at the same time take a so-called helicopter view on the whole chart. This is in contrast with classical process visualizations (i.e., process models) that mostly abstract from the data on individual cases and try to summarize the data. Both properties are beneficial for the explorative purpose the PPMCharts were developed for.

# 9 Acknowledgements

Our research builds upon the work of the development team of CEP and the researchers involved in the modeling sessions. Therefore, we express our extensive gratitude to Stefan Zugal, Jan Mendling and Dirk Fahland. We also thank the various people that provided feedback on the plug-in, subjects of the initial experiments and participants of the qualitative evaluation study. This research was funded by the Austrian Science Fund (FWF): P23699-N23.

[15] See http://software-document.blogspot.be/2010/07/activity-network-methods.html.

## APPENDIX A. Parameter settings of the PPMChart Analysis plug-in in ProM.

### A.1. Configuration

At the left hand side, the view can be configured (see Fig. 5 in the paper).

The ***Component type*** indicates which dimension is used to define the unit of the timelines. In contrast to the Dotted Chart Analysis plug-in, this option cannot be configured. The fixed value for this option in the PPMChart implementation is:

- *Model element* (default): select this option to view a *timeline per model element*. Each dot on the timeline represents an operation on the model element represented by the timeline (e.g., create, move, (re)name, delete of a particular XOR gateway).

The ***Time option*** can be configured to zoom in on the timing of the operations. Next three options can be selected:

- *Actual* (default): select this option to view the dots positioned according to the *real time* of execution of the corresponding operation.
- *Relative (Time)*: select this option to *shift* every time line in such a way that the first operation on each line is set to the beginning of the time interval of the PPMChart.
- *Relative(Ratio)*: select this option to *stretch* every timeline in such a way that the first operation on each line is set to the beginning of the time interval and the last operation on each line is set to the end of the time interval (if at least two operations exist on the line).

Vertical time intervals are marked according to the ***Time intervals*** configuration parameter. There are 13 different options.

- *L-1*, *L-10*, *L-100*, *L-500*: select these options to divide the chart in time intervals of 1, 10, 100, or 500 milliseconds respectively. Time intervals are indicated with white vertical lines starting at the time of the first operation in the chart. It is necessary to zoom in on the chart to be able to analyze the chart at millisecond level.
- *Seconds*, *Minutes*, *Half hours*, *Hours* (default): select these options to divide the chart in time intervals of seconds, minutes, half hours, or hours respectively.
- *Days*, *Week*, *Months*, *Years*: select these options to divide the chart in time intervals of days, weeks, months, or years respectively. It is necessary to zoom out on the chart to be able to analyze the chart at a level greater than one hour.

The option ***Color by*** indicates if the dots have to be color-coded or not. The PPMChart in principle uses a fixed default color coding (if turned on), but the colors can be changed by the user in the *Settings* tab (see A.3. below). Next two options can be selected:

- *None*: select this option to *remove color coding*. Each dot will have the same color, which allows the user to focus on shape and position of the dots (in order to abstract from the type of operation).
- *Operation* (default): select this option to *apply color coding*. By default, create operations will be colored in green, move operations in blue, delete operations in red, and (re)naming in orange. A detailed legend of the default colors is displayed in Table 3 below.

Use the ***Shape by*** setting to configure if the dots have to be shape-coded or not. The PPMChart in principle uses a fixed default shape coding (if turned on), but the shapes can be changed by the user in the *Settings* tab (see A.3. below). Next two options can be selected:

- *None*: select this option to *turn off dot shaping*. Each dot will be displayed as a circle, which allows the user to focus on color and position of dots (to abstract from the model element type of the operation).
- *Model element* (default): select this option to *turn on dot shaping*. Operations on activities will be displayed with rectangles, event operations with circles, gateway operations with diamonds, and edges with triangles. A detailed shape legend is displayed in Table 3 below.

***Sort by*** can be used to influence the order in which the timelines are sorted (vertically). If *descending* is selected, the sort order is reversed. Next eight options can be selected:

- *None*: select this option to select no ordering. The order of the data in the event log will be used.
- *Model element*: select this option to sort the lines by the *model element identifier*. The lines will be sorted according to the identifiers of the model elements represented by the timelines.
- *Number of operations*: select this option to sort the lines by the number of operations displayed on each line. Use this option to graphically observe differences between lines with fewer operations (top part of the chart if sorted according to this option) and lines with more operations (bottom part of the chart).
- *Duration*: select this option to sort the lines according to their duration. The duration is defined as the timespan between the first and the last operation on the line. This option allows to compare lines with shorter versus longer durations.

- *Distance from start* (default): select this option to sort the lines according to the traversing order of the corresponding model elements from the start event towards the end event (see description in Section 3.4.4 of the paper).
- *Create order from start*: select this option to sort the lines according to the logical order of creation of the corresponding elements from start event to end event (see description in Section 3.4.4 of the paper).
- *First operation*: select this option to sort the lines according to the time of the operation of the first dot on the line. This option facilitates to zoom in on the actual order of creation of model elements.
- *Last operation*: select this option to sort the lines according to the time of the operation represented by the last dot on the line. This option facilitates to zoom in on parts of the process model that are (not) touched towards the end of the modeling process.

Configure the ***Mouse mode*** to set the way the mouse behaves in the plug-in. Next three options can be selected:

- *Select* (default): select this option to be able to select different dots. Click on a dot or make a rectangular selection to indicate of which dots to display information in a tooltip.
- *Zoom in*: select this option to be able to easily zoom in on parts of the PPMChart. Make a rectangular selection on the screen to indicate the area you want to zoom in on.
- *Drag*: select this option to be able to bring a different area of the chart into the displayed rectangle if zoomed in. Drag the chart under the displayed rectangle to show other parts of the chart.

The sliders ***zoom (X)*** and ***zoom (Y)*** can be used to zoom in horizontal or vertical dimension respectively on a logarithmical scale. The ***Zoom out*** button restores the zoom level to 1 x 1. The ***Update*** button needs to be pressed after changing one or more of previous options before the PPMChart is repainted on the screen.

Table 3. Default color (shade) and shape coding of events

| | | Create | Move | Delete | (Re)name |
|---|---|---|---|---|---|
| | | Green | Blue | Red | Orange |
| **Start event** | Circle ● | CREATE_START_EVENT<br>Very light green circle | MOVE_START_EVENT<br>Very light blue circle | DELETE_START_EVENT<br>Very light red circle | NAME_ACTIVITY<br>RENAME_ACTIVITY<br>Orange square |
| **End event** | Circle ● | CREATE_END_EVENT<br>Very light green circle | MOVE_END_EVENT<br>Very light blue circle | DELETE_END_EVENT<br>Very light red circle | |
| **Activity** | Square ■ | CREATE_ACTIVITY<br>Green square | MOVE_ACTIVITY<br>Blue square | DELETE_ACTIVITY<br>Red square | |
| **XOR** | Diamond ◆ | CREATE_XOR<br>Dark green diamond | MOVE_XOR<br>Dark blue diamond | DELETE_XOR<br>Dark red diamond | |
| **AND** | Diamond ◆ | CREATE_AND<br>Dark green diamond | MOVE_AND<br>Dark blue diamond | DELETE_AND<br>Dark red diamond | |
| **Edge** | Triangle ▼ | CREATE_EDGE<br>Light green triangle<br><br>RECONNECT_EDGE<br>Light purple triangle | MOVE_EDGE_LABEL<br>Grey triangle<br><br>CREATE_EDGE_BENDPOINT<br>MOVE_EDGE_BENDPOINT<br>DELETE_EDGE_BENDPOINT<br>Dark grey triangle | DELETE_EDGE<br>Light red triangle<br><br>RECONNECT_EDGE<br>Light purple triangle | NAME_EDGE<br>RENAME_EGE<br>Orange triangle |

## A.2. Filtering

At the right-hand side the user can customize the view by filtering on specific operations or model elements (see Fig. 5 in the paper). The top part represents a small view on the unfiltered PPMChart. Below, one can configure next three filter options:

- *Hide next model elements*: choose to hide specific element types (e.g., hide edges).
  All dots that represent operations on an element of the selected type are removed from the chart. However, no timelines are removed. This might result in a PPMChart with a number of empty timelines (i.e., without any dot on the line).
- *Hide next operations*: choose to hide specific operation types (e.g., hide (re)name operations).
  All dots that represent operations of the selected types are removed from the chart. Again, only dots are removed from the chart, not timelines. Empty timelines may originate from this option if the model element represented by the timeline has only operations that are selected to be hidden.
- *Hide all elements with these operations*: hide elements with a specific operation (e.g., hide deleted elements). All dots that represent any operation on a model element that contains at least one operation of the selected operation type are removed from the chart. Again, only dots are removed from the chart, not timelines.

## A.3. Settings

Use the *Settings* tab page to change the color and shape coding of elements. Simply click on the button to change the color or shape for the corresponding operation.